\documentclass[11pt]{article}
\usepackage{jheppubmod,amsmath,amssymb}
\usepackage{color}
\usepackage{amsmath,amssymb}
\usepackage{comment}
\usepackage{braket}
\usepackage{mathtools}
\mathtoolsset{showonlyrefs}
\usepackage{psfrag}
\usepackage{array}
\usepackage{amssymb}
\usepackage{amsmath}
\usepackage{amsthm}
\usepackage{graphicx}
\usepackage{epstopdf}
	
\usepackage{color}
\usepackage{epsfig}
\usepackage[punctsep]{collref}
\collectsep[]{;}	

\def\vac{|0\rangle}
\def\co{{\cal O}}

\def\ta{\widetilde{a}}

\def\pb[#1,#2]{\{#1, #2\}}
\def\deb[#1,#2]{[#1,#2]_{\text{D.B.}}}

\def\tO{\widetilde{\cal O}}

\def\td{\widetilde{d}}

\def\ta{\widetilde{a}}

\def\Or[#1]{{\text{O}}\left({#1}\right)}
\def\dotl[#1,#2]{\left\langle #1,\, #2 \right\rangle}
\def\dotlb[#1,#2]{\left\langle #1,\, #2 \right\rangle}
\def\dotlm[#1,#2]{\left[ #1,\, #2 \right]}
\def\dotp[#1,#2]{(\vect{#1} \cdot\vect{#2})}
\def\aff[#1,#2]{\hat{#1}(#2)}

\def\n4sym{{\cal N}=4 SYM}
\def\>{\rangle}
\def\<{\langle}
\def\weight[#1,#2,#3]{\{(#1),#2,#3\}}
\def\ads[#1]{$\text{AdS}_{#1}$}

\def\tarelr[#1]{\widetilde{a}^{\text{rel}}_{R#1}}
\def\Oright[#1]{{\cal O}_{R#1}}
\def\Oleft[#1]{{\cal O}_{L#1}}
\def\aleft[#1]{a_{L#1}}
\def\arelr[#1]{a^{\text{rel}}_{R#1}}

\newcommand{\be}{\begin{equation}}
\newcommand{\ee}{\end{equation}}
\newcommand{\ba}{\begin{align}}
\newcommand{\ea}{\end{align}}
\newcommand{\bs}{\begin{split}}
\def\sess\end{split}

\newcommand{\vect}[1]{{#1}}

\def\tO{\widetilde{\cal O}}
\def\rs{|\Psi\rangle}
\def\ls{\langle \Psi|}
\def\rsp{|\Psi'\rangle}
\def\lsp{\langle \Psi'|}
\def\rsz{|\Psi_{0}\rangle}
\def\lsz{\langle \Psi_0|}

\def\Tr{{\rm Tr}}
\def\tO{\widetilde{\cal O}}
\def\rs{|\Psi\rangle}
\def\ls{\langle \Psi|}
\def\rsp{|\Psi'\rangle}
\def\lsp{\langle \Psi'|}
\def\cS{{\cal S}}

\def\pP{\mathbb P}

\author{Kyriakos Papadodimas}
\emailAdd{kyriakos.papadodimas@cern.ch}
\affiliation{Theoretical Physics Department, CERN, CH-1211 Geneva 23,
Switzerland}
\affiliation{Van Swinderen Institute for Particle Physics and Gravity, University of Groningen, Nijenborgh 4,
9747 AG Groningen, The Netherlands}
\abstract{We consider a class of non-equilibrium pure states, which are generally present in an isolated quantum statistical system. These are states of the form $\rs=e^{-{\beta H \over 2}} U({\cal O}) e^{{\beta H \over 2}} \rsz$, where $U$ is a unitary made out of simple operators and $\rsz$ is a typical equilibrium pure state with sharply peaked energy. We argue that in a system with a holographic dual these states have a natural interpretation as an AdS black hole with transient excitations behind the horizon.
We explore the  interpretation of these states as pure states undergoing a time-dependent spontaneous fluctuation out of equilibrium. While these states are atypical and the microscopic phases of the wavefunction are correlated with the matrix elements of simple operators,  the states are partly disguised as equilibrium states due to cancellations between contributions from different coarse-grained energy bins. These cancellations are guaranteed by the KMS condition of the underlying equilibrium state $\rsz$. However, in correlators which include the Hamiltonian $H$ these cancellations are spoiled and the non-equilibrium nature of the state $\rs$ can be detected. We discuss connections with the proposal that local observables behind the horizon are realized as state-dependent operators. The states studied in this paper may be useful for implementing an analogue of the ``traversable wormhole'' protocol for a 1-sided black hole, which could potentially allow us to extract the excitation from behind the horizon. We include some pedagogical background material.

}
\preprint{\\\hspace*{\fill}
CERN-TH-2017-160	
\vspace{-30pt}
}

\begin{document}
\title{\Large A class of non-equilibrium states and the black hole interior \vspace{-15pt}}

\maketitle

\newpage

\section{Introduction}

In this paper we investigate a class of non-equilibrium states in quantum statistical mechanics and their gravitational interpretation via AdS/CFT \cite{Maldacena:1997re,
Gubser:1998bc,Witten:1998qj} as states with excitations localized behind the black hole horizon. 
These are states  of the form
\be
\label{introstatesa}
|\Psi\rangle = e^{-{\beta H \over 2}} U({\cal O}) e^{\beta H \over 2}\rsz
\ee
where $\rsz$ is a typical pure state of energy $E_0$, $\beta$ is the inverse temperature corresponding to this energy and $U$ is a unitary operator constructed out of ``simple operators'' ${\cal O}$ centered around some time $t_0$. 
The states \eqref{introstatesa} have the property that they seem to be in equilibrium when probed by a class of other simple operators,  up to $1/S$ corrections where $S$ is the entropy of the system. Nevertheless they are genuinely time-dependent, non-equilibrium, atypical states.

These states are interesting for two reasons. First, they represent a mathematically natural class of non-equilibrium states in general thermal systems, which have not been discussed extensively in the literature. Second, if we accept that big black holes in AdS have a smooth interior, then we argue that states of the form \eqref{introstatesa} correspond to  black holes with excitations localized entirely behind the horizon. Investigating the properties of these states may provide insights about the nature of space-time behind the black hole horizon.

The recent reformulation of the black hole information paradox in terms of the entanglement of quantum fields near the horizon has led to proposals \cite{Mathur:2009hf,Almheiri:2012rt} that the black hole interior may be different than what is predicted by general relativity. In particular, it has been  argued \cite{Almheiri:2013hfa, Marolf:2013dba} that big black holes in AdS may not have a smooth interior\footnote{Strictly speaking the arguments of these papers apply to typical microstates.}, or that even if they do, the boundary CFT cannot describe it. 
In a series of works \cite{Papadodimas:2012aq,Papadodimas:2013b,Papadodimas:2013,Papadodimas:2013kwa,Papadodimas:2015xma,Papadodimas:2015jra} 
a construction of the black hole interior within AdS/CFT was proposed. A novel feature of this proposal is that, depending on the black hole microstate,  local observables behind the horizon are represented by different linear operators acting on certain subspaces of the Hilbert space. This property, termed state-dependence, has also manifested itself in somewhat different form in \cite{Verlinde:2012cy,Verlinde:2013uja} and in the ER/EPR correspondence \cite{Maldacena:2013xja}. See  \cite{Almheiri:2013hfa, Marolf:2013dba, luboscoherentblog, Harlow:2014yoa,Marolf:2015dia,Banerjee:2016mhh,Raju:2016vsu,Berenstein:2016pcx,Berenstein:2017abm,Jafferis:2017tiu,Marolf:2017jkr,Raju:2017ost,Berenstein:2017rrx} for further discussions.

In this paper we do not focus on the {\it operators} but rather on {\it states} in which the black hole interior is excited. Some of these states can be represented as \eqref{introstatesa}, where we 	 see that they  can be constructed in the CFT without the need for state-dependent operators. The existence of the states \eqref{introstatesa} in the standard framework of quantum mechanics is obvious. We mention that constructing the states  does not supersede the need to define  operators, since in order to describe a quantum measurement for the infalling observer we also need  the operators and not just the states. More precisely, if we want to know that we have the correct bulk physical interpretation of the states \eqref{introstatesa}, we need to know how bulk observables probe the state. Nevertheless, the fact that there exists a mathematically canonical class of states \eqref{introstatesa} in the CFT which, as we will argue, can be naturally interpreted as states with excitations behind the horizon can be taken as additional evidence that the black hole interior can be described in the CFT.

An equilibrium state in a strongly coupled large $N$ CFT is holographically dual to a static\footnote{We only consider equilibrium states with vanishing angular momentum and other conserved global charges.} black hole in AdS. This implies that there should be a correspondence between small perturbations of the state on the two sides of the duality: the set of non-equilibrium states, which are small perturbations of the original equilibrium state, should be related to the possible excitations of the solution in the bulk, which  depend on the form of the underlying geometry.
Hence, classifying the possible non-equilibrium states\footnote{In this paper we consider near-equilibrium states where the perturbation changes the energy of the state by an amount of order $N^0$, so the backreaction to the classical bulk metric is negligible.} in the CFT contains information about the geometry in the bulk. We do not attempt a full classification, but we concentrate on a particular family of non-equilibrium states \eqref{introstatesa} which correspond to excitations localized entirely behind the black hole horizon.

In general, we can think of an equilibrium pure state $\rsz$ as a typical state, i.e. a state selected with the unitary Haar measure from the states of given energy.
One way to consider non-equilibrium states is to start with an equilibrium state $\rsz$ and excite it by acting with a unitary operator $U({\cal O}(t_0))$, which is expressed in terms of some ``simple'' operators ${\cal O}$. In a large $N$ CFT these could be smeared single-trace operators of low conformal dimension. This gives the state $U({\cal O}(t_0))\rsz$, which is out of equilibrium 
around the time $t_0$, and which settles back to equilibrium for $t\gg t_0$. One way to interpret this state is as the result of a quench of the system, where we add to the Hamiltonian a source constructed out of ${\cal O}$  around $t\approx t_0$. Alternatively we can think of the state $U({\cal O}(t_0))\rsz$
as an ``autonomous state'', where the system evolves with the standard CFT Hamiltonian for all times, with no external perturbation. From this second point of view this is a state which is in equilibrium for $t\ll t_0$, undergoes a spontaneous fluctuation out of equilibrium around the time $t = t_0$ and then settles back to equilibrium for $t\gg t_0$.

Similarly, the  non-equilibrium states of the form
$
 e^{-{\beta H \over 2}} U({\cal O}(t_0)) e^{\beta H \over 2}\rsz
$ can be thought of as autonomous states which are in equilibrium for $t\ll t_0$ and $t\gg t_0$, but which undergo a spontaneous fluctuation out of equilibrium at $t= t_0$. The difference with states of the form  $U({\cal O}(t_0))\rsz$  is that now the excitation at $t=t_0$ is not directly detectable by the appropriately defined algebra of simple operators. We emphasize that, for reasons that will become clear later, the Hamiltonian $H$ is not included in this algebra.

\begin{figure}[!t]
\label{fig1}
\begin{center}
\includegraphics[width=.8\textwidth]{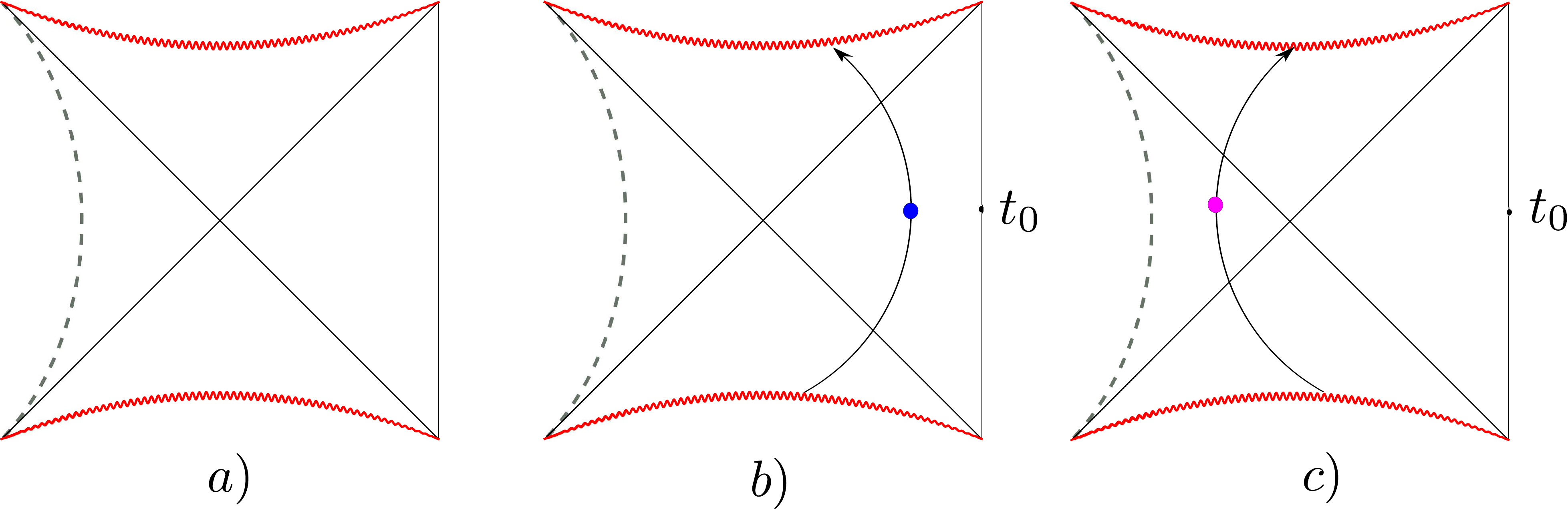}
\caption{ a) Typical equilibrium state $\rsz$ b) Non-equilibrium state of the form $U({\cal O})\rsz$ with an excitation in free-fall c) Non-equilibrium state of the form $e^{-{\beta H \over 2}} U({\cal O}) e^{\beta H \over 2}\rsz$ with a similar excitation. Notice that in these figures we show the conjectured Penrose diagram
for a {\it one-sided} black hole and the dual CFT corresponds to the right asymptotic region. There is no dual CFT on the left side of the diagram and it is not clear until what distance the diagram can be continued into the left region.}
\end{center}
\end{figure}

If we are dealing with a statistical system with a gravitational dual, for example a strongly coupled large $N$ CFT, then the proposed bulk interpretation of the various  states mentioned above is shown in figure \ref{fig1}. Typical equilibrium states $\rsz$ correspond to a static black hole
geometry. The usual non-equilibrium states $U({\cal O}) \rsz$, seen as autonomous states,	 correspond to spontaneous fluctuations emitted from the past horizon and falling into the future horizon. Assuming that the horizons are smooth, these excitations seem to appear from the past singularity and fall back into the future singularity.  Finally the states \eqref{introstatesa} correspond to similar excitations emitted from the past singularity, but which are now ejected towards the left region of the diagram and eventually fall back into the future singularity. The part of the spacetime to the right of the past and future horizons can be 
reconstructed by considering correlators of standard CFT operators. Using the state-dependent operator construction of \cite{Papadodimas:2013} on the states \eqref{introstatesa} we do indeed get the proposed bulk interpretation in figure \ref{fig1}.

Typical states $\rsz$ look like equilibrium states because when they are expanded in a basis of energy eigenstates the coefficients of the superposition have uncorrelated phases with the matrix elements of simple observables. This can also be understood from the Eigenstate Thermalization Hypothesis (ETH) \cite{peresETH,deutsch,srednicki1999approach}. Usual non-equilibrium states of the form $U({\cal O})\rsz$ have the property that these phases become correlated with matrix elements of certain simple observables, which can be used to detect the excitation away from equilibrium. In states of the form $ e^{-{\beta H \over 2}} U({\cal O}) e^{\beta H \over 2}\rsz$ the phases are  as correlated as in the states $U({\cal O})\rsz$, however it is difficult to detect the non-equilibrium nature of the state because there are cancellations between ``energy bins'' of higher and lower energies. Conjugating the perturbation $U({\cal O})$ with the factors $e^{\pm{\beta H \over 2}}$ has the effect of enhancing the lower energy bins and suppressing the higher energy bins, which in combination with the KMS condition of the underlying equilibrium state $\rsz$ leads to the aforementioned cancellations. This means that, even though both typical states $\rsz$ and  states of the form $ e^{-{\beta H \over 2}} U({\cal O}) e^{\beta H \over 2}\rsz$ give time-independent correlators for simple operators (excluding the Hamiltonian $H$), the reason for the time-independence is qualitatively different. In the first case time-independence is the result of randomized phases, while in the second case the phases are not random, but nevertheless the cancellations happen because  of special properties of the {\it magnitudes} of the coefficients.

Inserting the Hamiltonian $H$ inside the correlator can have the effect of spoiling these cancellations, leading to time-dependent correlators for the states $ e^{-{\beta H \over 2}} U({\cal O}) e^{\beta H \over 2}\rsz$. This allows us to distinguish these states from ``true equilibrium states'' in which all correlators of simple operators, now including $H$, are time-independent. We  summarize the properties of the states discussed so far in table \ref{tab1}.

\begin{table}
\begin{center}
    \begin{tabular}{| m{1.5cm}|  m{4cm}|m{4cm}| m{4cm}|}
    \hline
     & True equilibrium state  & Standard non-eq. state & ``Interior'' non-eq. state  \\ \hline
     & $\qquad  \quad \,\,\,\,\,\rsz$  & $\qquad \quad U({\cal O})\rsz$ & $\quad e^{-{\beta H \over 2}}U({\cal O}) e^{{\beta H \over 2}}\rsz$ \\ \hline
    ${d\over dt}\langle A(t) \rangle$  & $\qquad\qquad$ $0$ & $\qquad\qquad$ $\neq 0$ & $\qquad\qquad\,\,$  $0$ \\ \hline
    ${d\over dt}\langle A(t) H\rangle$ &$\qquad\qquad$  $0$ & $\qquad\qquad$  $\neq 0$ & $\qquad \quad\,\,\,\,$ $\neq 0$ \\
    \hline
    \end{tabular}
    \end{center}
    \caption{We summarize properties of the various states discussed in this paper. We define the ``small algebra'' ${\cal A}$ of simple operators, in which we {\it do not} include the Hamiltonian $H$. Here $U({\cal O})$ is a simple unitary constructed out
    of operators ${\cal O}$ in ${\cal A}$. By $A(t)$ we denote a general operator in ${\cal A}$. The notation $\langle...\rangle$ means expectation value in the corresponding state.
    The statements about the time-dependence refer to the result at order $S^0$. When we say that the time-derivative is $\neq 0$ we mean that $A(t)$ may be selected so that the correlator is time-dependent. The state $\rsz$ denotes a typical pure state with small spread in energy.}
    \label{tab1}

    \end{table}

The existence and certain properties of the states \eqref{introstatesa} are logically independent from the state-dependent proposal \cite{Papadodimas:2013}  and most of the paper can be read independently. The fact that we do not need state-dependence for the construction of the states, as opposed to the operators, should not be surprising. For example, even in non-gravitational QFT in Minkowski space, the Reeh-Schlieder theorem  \cite{Reeh1961} implies that we can effectively construct the full Hilbert space of states by acting with operators in the right Rindler wedge. However, it is not possible to express the  operators of the left wedge in terms of operators on the right wedge. Notice that the same distinction between creating states as opposed to operators also applies to the recent work \cite{Goto:2017olq}. 

The black holes that we consider are small perturbations of typical states and in that sense we are very far from the regime relevant for a black hole formed by gravitational collapse, which corresponds to special, atypical states. In a collapsing black hole states with excitations behind the horizon with some similarities  to \eqref{introstatesa} can be created by sending particles from the exterior, before the collapse. This is discussed in more detail in subsection
\ref{seccomments}. It would be interesting to understand better how to interpolate between the two regimes of typical vs collapsing black holes.

Some aspects of states of the form \eqref{introstatesa} have been considered previously in \cite{Papadodimas:2013, Harlow:2014yoa} in connection to the firewall paradox and proposals for its resolution. In this paper we want to investigate certain properties of these states from the point of view of statistical mechanics.

It would be interesting to find additional support for the bulk interpretation of the states $e^{-{\beta H \over 2} } U({\cal O}) e^{{\beta H \over 2}} \rsz$ proposed in this paper. In recent work \cite{Gao:2016bin,Maldacena:2017axo} it was shown that there are double-trace perturbations of the TFD state of two entangled CFTs, 
which allow excitations in the bulk to cross the wormhole of the eternal AdS black hole. This happens by creating a negative energy shockwave which gives the probe a time-advance necessary to escape from the horizon. In the 2-sided case the double-trace perturbation can also be related to a quantum teleportation protocol between the two CFTs. These works have provided evidence for the smoothness of the horizon of the eternal black hole.

\begin{figure}[!t]
\begin{center}
\includegraphics[width=.6\textwidth]{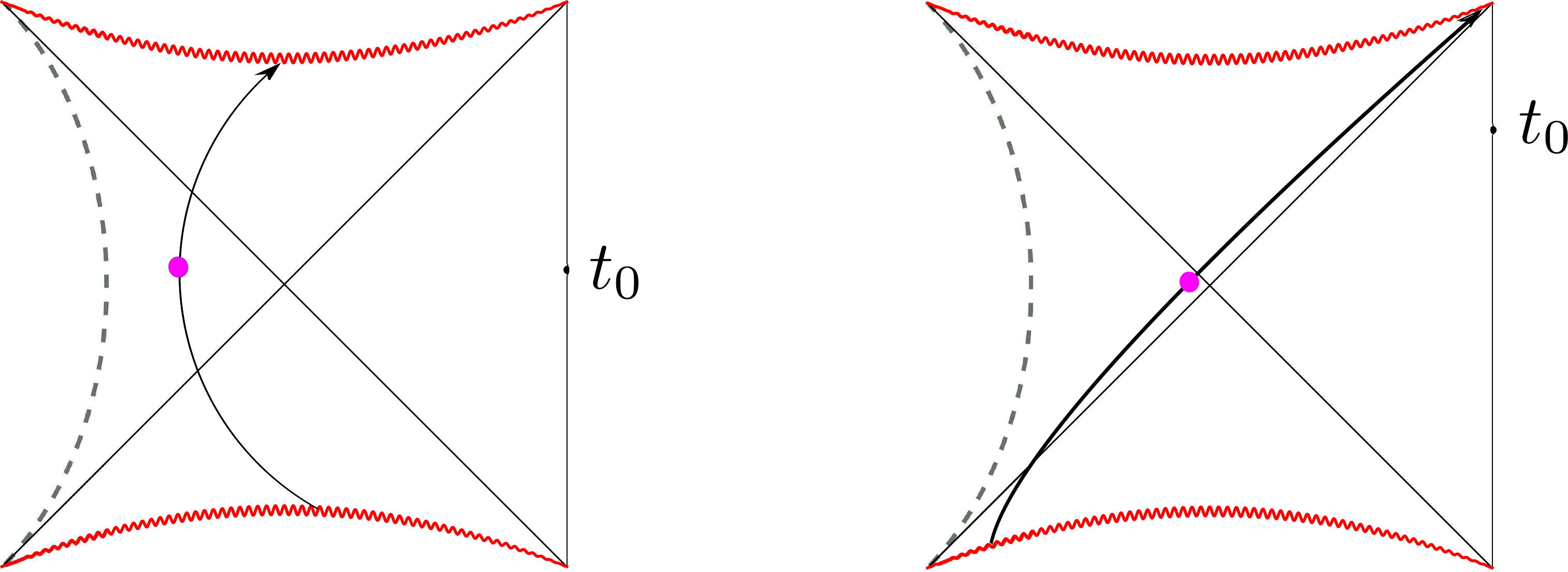}
\caption{Left: a state of the form $e^{-{\beta H \over 2} } U({\cal O}) e^{{\beta H \over 2}} \rsz$. Right: the same state seen at an earlier time.}
\label{boostp1}
\end{center}
\end{figure}

A natural question is whether a similar experiment can be designed for the 1-sided black hole. One difficulty is preparing the probe that is supposed to be extracted via the negative energy shockwave. The states of the form $\rs=e^{-{\beta H \over 2} } U({\cal O}) e^{{\beta H \over 2}} \rsz$ provide a natural starting point since they already contain a particle behind the horizon. In figure \ref{boostp1} we see the state $\rs$ and how it looks after a time translation. We notice the right-moving particle traveling close to the future horizon. A negative energy shockwave emitted from the boundary CFT living on the right might be able to displace the particle and allow it to escape the black hole, thus allowing it to be directly detected in the CFT by usual single trace operators ${\cal O}$. See \cite{Kourkoulou:2017zaj} for related discussions in the context of the SYK model.  If such a calculation was possible, it would provide additional support for the proposed interpretation of the states \eqref{introstatesa} discussed in this paper, as well as for the proposal of \cite{Papadodimas:2012aq,Papadodimas:2013b,Papadodimas:2013,Papadodimas:2013kwa,Papadodimas:2015xma,Papadodimas:2015jra} for operators which can directly detect the excitation behind the horizon. We leave this for future work.

A question that we do not address in this paper is to identify the precise bulk interpretation of states related to unitaries $U({\cal O})$ which approximately commute with the Hamiltonian $H$. These were discussed in \cite{Almheiri:2013hfa,Harlow:2014yoa,Papadodimas:2015jra,Marolf:2015dia,Raju:2016vsu,Raju:2017ost}. Clarifying the role of these perturbations is necessary for a full mapping between all possible non-equilibrium states and excitations of the black hole geometry. It would also be necessary to understand the full class of perturbations including unitaries on both sides of the horizon of a 1-sided black hole.

The plan of the paper is as follows: in section 2 we discuss aspects of thermalization in isolated quantum systems and how a second copy of the operator algebra can be defined in a canonical way. In section 3 we present some properties of localized states in flat space QFT in Rindler coordinates and we notice some similarities when we move an excitation to the other side of a Rindler horizon using a complexified Lorentz boost. In section 4 we briefly review results 
from quantum statistical mechanics. A more extended discussion of this material is presented in appendix A. In section 5 we analyze equilibrium and a class of non-equilibrium states and discuss their bulk interpretation. In section 6 we discuss some additional statistical-mechanics aspects of the states and in section 7 we end with some discussions.

\section{Thermalization, small algebras and doubling}
\label{sectherm}

We  expect that an isolated quantum system with many degrees of freedom and strong interactions will generally approach thermal equilibrium under time evolution, even if it is not coupled to an external heat bath. 
This means that if we take the system in some initial pure state $\rsz$  whose average energy is $E_0$ and wait until the system equilibrates, then for a class of observables ${\cal O}$ we will have
\be
\label{thermalizationa}
\lsz {\cal O}(t) \rsz\rangle = Z^{-1}{\rm Tr}[e^{-\beta H}{\cal O}(t)] + O(S^{-1})
\ee
Here $S$ is the entropy, $\beta$ is the inverse temperature associated to the energy $E_0$ of the state $\rsz$ via $\beta = {\partial S \over \partial E}$ and $Z \equiv {\rm Tr}[e^{-\beta H}]$. When estimating the error term\footnote{The error term is of order $1/S$
and not of order $e^{-S/2}$ because we are comparing a pure state with sharply peaked energy to the canonical (instead of the microcanonical) ensemble. More about it in section 4.}  we assumed that the operator ${\cal O}$ has been scaled so that 
its norm scales with entropy like $S^0$.

We expect equation \eqref{thermalizationa} to hold for certain observables ${\cal O}$, that we will call the ``small algebra ${\cal A}$''. The restriction to a small algebra is natural in the study of thermalization, since in order to talk about thermalization of an isolated system in a pure state, we must assume that we are not able to measure everything --- otherwise we would be able to measure very complicated operators which do not obey \eqref{thermalizationa} and which allow us to 
identify the exact pure state. This requires that there must be a hierarchy between the total entropy of the system $S$ and the size of the small algebra ${\cal A}$, which we define by
starting with some ``simple'' operators and considering their products, imposing the condition that the number of factors $n$ in the product should obey $n\ll S$. Because of this cutoff
the set ${\cal A}$ is not really an algebra. However in the large $S$ limit this cutoff is not important for many purposes and we will usually refer to the set ${\cal A}$ as an algebra. In a large $N$ gauge theory the small algebra is generated by ``simple'' $SU(N)$ single trace operators.

The restriction to a small algebra allows us to effectively consider the isolated system as being made out of a subsystem, corresponding to ${\cal A}$, and an environment. In this sense, part of the isolated system 
plays the role of the heat bath for the small algebra ${\cal A}$. In the systems that we are interested in the small algebra does not correspond to a subregion in space, but rather to a set of ``simple operators'' which can be placed anywhere in space\footnote{We are considering systems defined in finite spatial volume.}. So the decomposition in terms of subsystem and effective environment is not spatial, but rather characterized by the complexity of the observables.

Equation \eqref{thermalizationa} implies that a system in a pure state $\rsz$ can be approximated by the thermal density matrix $\rho_\beta = Z^{-1} e^{-\beta H}$. A system in a thermal density matrix can also be described by a pure state in a doubled Hilbert space in the ``thermofield-doubling'' framework
\be
\label{tfdform}
|\Psi_{\rm tfd}\rangle = \sum_E {e^{-\beta E/2}\over \sqrt{Z}} |E\rangle \otimes |E\rangle
\ee
Computing the reduced density matrix of the original system on this entangled state, we get the thermal density matrix. In this framework, where we consider the purification of the thermal density matrix by an auxiliary second copy of the system, we notice that the algebra of observables is doubled. However, the physical meaning of the doubling of the algebra is not obvious. In addition, equation \eqref{thermalizationa} suggests that a similar construction should be possible, at least in an approximate sense, in the case where the original system is in a pure state $\rsz$, and not in a thermal density matrix. But then, if the isolated system is already in a pure state to begin with, it seems like it would not make sense to purify it using a second copy.

In \cite{Papadodimas:2012aq,Papadodimas:2013b,Papadodimas:2013} a proposal was made which addresses these points. The original motivation was to understand how to describe the black hole interior in the framework of AdS/CFT, but the interpretation of the thermofield-double formalism proposed in those papers could apply to more general quantum statistical systems. An intuitive motivation of the proposal of \cite{Papadodimas:2012aq,Papadodimas:2013b,Papadodimas:2013}  is that the small subsystem is "mirrored" in the effective heat bath (which is also part of the original system). This means that there is certain class of excitations of the heat bath
which mirror the dynamics of the small subsystem. These excitations are selected by their entanglement with the small subsystem.  In the case of a single-sided black hole in AdS, the small subsystem is the exterior of the horizon, as described in effective field theory, which is mirrored in the region behind the horizon.

The starting point is to consider the algebraic properties of the representation of the algebra ${\cal A}$ associated to the state $\rsz$. If the state $\rsz$ is a highly excited pure state, for which we can use the approximation \eqref{thermalizationa},  then the small algebra ${\cal A}$ contains no (non-vanishing) annihilation operators for
the state $\rsz$. For example, the norm-square of the state $A \rsz$ can be written as $\lsz A^\dagger A \rsz = Z^{-1} {\rm Tr}[e^{-\beta H} A^\dagger A] + O(1/S)$. If the operator $A$ is not identically zero, then to leading order in $S$ we have ${\rm Tr}[e^{-\beta H} A^\dagger A] >0$, since $e^{-\beta H}$ is positive definite. As we will see below, the absence of annihilation operators in the algebra ${\cal A}$ for the state $\rsz$ means that ${\cal A}$ probes the state $\rsz$ as if it were an entangled state. The  auxiliary second copy in the thermofield formalism \eqref{tfdform} is interpreted as a way to represent this effective entanglement of the state $\rsz$ with respect to the algebra ${\cal A}$.

A technical point is that when we define the algebra ${\cal A}$, we do not include in it the Hamiltonian $H$. The reason is that we are working with states $\rsz$ with average energy $E_0$ and small uncertainty in energy. Hence these states obey $(H - E_0) \rsz \approx 0$. This would allow us to find an approximate annihilation operator in the algebra ${\cal A}$, which would complicate the following algebraic construction. For this reason we exclude $H$ from the small algebra ${\cal A}$. However, if $A \in {\cal A}$ then the operator $[H,A]$ is contained in ${\cal A}$. Hence, ${\cal A}$ is closed under time evolution, at least for small time intervals. While $[H,A]$ is in the algebra, operators like $AH$ or $\{H,A\}$ are not included. Similar statements hold for other conserved charges, see \cite{Papadodimas:2013} for more details.

We continue by reviewing the construction of \cite{Papadodimas:2012aq,Papadodimas:2013b,Papadodimas:2013}. We introduce a subspace of the full Hilbert space
\be
\label{smallh}
{\cal H}_{\Psi_0} = \text{span}\,\{ {\cal A} \rsz \,\}
\ee
This is similar to what was called the ``code-subspace'' in later work \cite{Almheiri:2014lwa}. 

We now concentrate on the representation of the algebra ${\cal A}$ on the space ${\cal H}_{\Psi_0}$. This representation satisfies two properties. First, by definition the   space ${\cal H}_{\Psi_0}$ can be generated by acting with elements of the algebra ${\cal A}$ on the vector $\rsz$. Second,
there are no operators in the algebra ${\cal A}$ which can annihilate the state $\rsz$. These two conditions are sometimes called respectively, {\it cyclic} and {\it separating} properties. 

These conditions are similar to the assumptions of the Tomita-Takesaki theorem in the study of operator algebras, see for example \cite{bropalg,Haag:1992hx}. This theorem starts with the assumption that a von Neumann algebra ${\cal A}$ acts on a Hilbert space with a cyclic and separating vector $\rsz$. The theorem 
shows that the representation of the algebra ${\cal A}$ on the Hilbert space is reducible and the algebra has a non-trivial commutant ${\cal A}'$, which is isomorphic to the original algebra and which acts on the same space\footnote{More precisely, in general  the algebras ${\cal A}$ and ${\cal A}'$ have domains $D,D'$ which are dense in the Hilbert space.}. Moreover it shows that there is a canonical 1-parameter automorphism of the algebra generated by a distinguished Hermitian operator called the modular Hamiltonian. The commutant ${\cal A}'$ plays a similar role as the algebra of operators acting on the second
auxiliary copy of the thermofield construction \eqref{tfdform}. We emphasize that in the Tomita-Takesaki framework the Hilbert space is not doubled: the commutant of the algebra is acting on the original Hilbert space.

The Tomita-Takesaki theorem deals with von Neumann algebras. In the situation of an isolated bounded quantum system we introduced the ``small algebra'' ${\cal A}$ where we did not allow the multiplication of arbitrary number of operators. Hence  ${\cal A}$ is not really a proper algebra, but rather a ``truncated algebra'' defined by some generators which can be multiplied up to an upper bound on the number of factors. In a large $N$ gauge theory the generators can be taken to be the single trace operators\footnote{Another complication is that the Tomita-Takesaki theorem applies to algebras of bounded operators. It is generally non-trivial to define bounded operators in QFT which are localized in finite regions of spacetime.}, and the bound is that when multiplying them together we do not allow the number of factors to scale like $N$. The fact that we impose this upper cutoff by truncating the algebra ${\cal A}$ means that the Tomita-Takesaki theorem cannot by applied in an exact sense for the bounded quantum system in a pure state. Relatedly,  if we enlarge the small algebra ${\cal A}$ to include arbitrary products, thus promoting it into a proper algebra, then we {\it can} find annihilation operators for the state $\rsz$ hence the separating property breaks down and then there is no non-trivial exact commutant\footnote{This is consistent with the time-slice axiom in QFT.}. In applications to black hole physics, this fact is important as it leads to a realization of the idea of black hole complementarity \cite{'tHooft:1990fr,Susskind:1993if} as explained in \cite{Papadodimas:2012aq,Papadodimas:2013b,Papadodimas:2013}. This will not play a role in the rest of this paper. 

Despite the fact that it is subtle to apply the theorem in an exact mathematical sense, we expect that in the large $N$ limit the intuition behind the Tomita-Takesaki construction should become relevant for the states within ${\cal H}_{\Psi_0}$. Thus we are going to follow the Tomita-Takesaki construction from a heuristic point of view. Our eventual goal is to motivate equations \eqref{todefa}, which can in any case be taken as the definition of the second copy of the algebra.

We proceed describing the steps, as if the algebra ${\cal A}$ were a von Neumann algebra.
The Tomita-Takesaki theorem gives a canonical construction of the commutant ${\cal A}'$ of the algebra ${\cal A}$. We start by introducing the antilinear map $\cS: {\cal H}_{\Psi_0} \rightarrow {\cal H}_{\Psi_0}$ defined by\footnote{Generally the domain of $\cS$ may be a dense subspace of the  Hilbert space.}
\be
\label{smap}
\cS {\cal O} \rsz \equiv {\cal O}^\dagger \rsz \qquad {\cal O}\in {\cal A}
\ee
That this mapping can be consistently defined, relies on the fact that the algebra ${\cal A}$ has no annihilation operators for the state $\rsz$. Otherwise we may be able to find an operator with the property ${\cal O} \rsz = 0$, while ${\cal O}^\dagger \rsz \neq 0$, which would make inconsistent the defining equation \eqref{smap} for the anti-linear operator $\cS$. This is precisely the reason why the doubling does not work for ---say--- the ground state of the system, for which many annihilation operators can be found. 

We  proceed by considering the polar decomposition 
\be
\label{defj}
{\cal S} =J\Delta^{1/2}
\ee
where $J$ is an anti-unitary operator and 
\be
\label{moddef}
\Delta \equiv \cS^\dagger \cS
\ee

Since $\Delta$ is a positive Hermitian operator\footnote{For infinite dimensional algebras, the operator $\Delta$ may be unbounded
and we have to be careful about its domain. For more details see \cite{bropalg}.}, we can write it as $$\Delta = e^{-K}$$ The exponent $K\equiv -\log(\cS^\dagger \cS)$ is defined as the modular Hamiltonian for the algebra ${\cal A}$ acting on the state $\rsz$.

Finally from the Tomita-Takesaki theorem it follows that for any ${\cal O}\in {\cal A}$ the operator $\tO$ defined as
\be
\label{defto}
\tO \equiv J {\cal O} J
\ee
commutes with all elements $A$ of the algebra ${\cal A}$
\be
\label{comA}
[A,\tO] = 0
\ee
and we have ${\cal A}' = J {\cal A} J$. This means that the entire algebra ${\cal A}$ and its commutant ${\cal A}'$ are isomorphic via the antilinear map $J$. We also mention that another important part of the theorem is that the algebra ${\cal A}$ is invariant under the modular flow
$$
e^{it K} {\cal A} e^{-i t K}  = {\cal A} \quad \forall t\in {\mathbb R}
$$
For the proof of these statements see \cite{bropalg}.

However notice that generally
\be
\label{comK}
[K,{\cal O}]\neq 0 \quad,\quad [K,\tO]\neq 0.
\ee
so the modular Hamiltonian is neither an element of the algebra ${\cal A}$ nor of the commutant ${\cal A'}$, but rather it has support on both. If we apply the Tomita-Takesaki construction to the algebra of a subsystem of a bipartite system in a pure state, then the modular Hamiltonian, as defined by \eqref{moddef}, is the same as the full modular Hamiltonian defined via the logarithm of the reduced density matrices of the
subsystem and its complement, see for example \cite{Papadodimas:2013}. In the Rindler decomposition of the Minkowski vacuum the modular Hamiltonian \eqref{moddef} defined for the right Rindler wedge algebra, coincides with the full Lorentz boost generator, and not its restriction on the right wedge, as we review in the next section.

So far the discussion was general, we only used the assumption that $\rsz$ is a cyclic and separating vector with respect to the algebra ${\cal A}$. To apply these steps to a typical equilibrium pure state $\rsz$ of a quantum statistical system with Hamiltonian $H$ we need to concentrate on the part of the space ${\cal H}_{\Psi_0}$ and the operators in the algebra ${\cal A}$ which are ``far'' from the cutoff of the algebra\footnote{In \cite{Papadodimas:2013} the (in)sensitivity to the cutoff was discussed, but it would be useful to consider it in more detail.}. Using the fact that $\rsz$ is an equilibrium state which obeys the KMS condition to leading order in $1/N$ then we find that\footnote{The precise statement is that the operator
$\beta (H-E_0)$ and $K$ act in the same way, at large $N$, {\it within the subspace} ${\cal H}_{\Psi_0}$, see sec. 6.2 of \cite{Papadodimas:2013} for a proof.}
\be
\label{modeq}
K=- \log(\cS^\dagger \cS) = \beta (H-E_0) + O (1/N)
\ee
where $E_0$ is the expectation value of the energy of the state $\rsz$. For this result it is important that $\rsz$ is a state with small spread in energy.

This allows us to rewrite the $\widetilde{\cal O}$ operators more explicitly. For every element ${\cal O}$ of the algebra ${\cal A}$ satisfying the rules of the large $N$ expansion (in a large $N$ gauge theory, for every single trace operator), we define
the Fourier modes in time ${\cal O}_\omega\equiv \int dt e^{i \omega t} {\cal O}(t)$. Then the equations \eqref{smap}, \eqref{moddef}, \eqref{defto} above imply that to leading order in $1/N$ we have
\begin{align}
\label{todefa}
& \widetilde{\cal O}_\omega A \rsz =  A e^{-{\beta \omega \over 2}} {\cal O}^\dagger_\omega \rsz \cr
& [H,\widetilde{\cal O}_\omega ] A \rsz =  \omega \widetilde{\cal O}_\omega  A \rsz
\end{align}
for all elements $A$ of the algebra ${\cal A}$. Since any vector from the subspace \eqref{smallh} can be written as $A\rsz$ for some $A\in {\cal A}$, these equations fully define the tilde-operators on the subspace \eqref{smallh}. More details about this construction can be found in  \cite{Papadodimas:2013}.

A point to stress is that these equations define the operators $\widetilde{\cal O}$ acting on the subspace ${\cal H}_{\Psi_0}$ associated to the microstate $\rsz$. If we consider a different microstate $|\Psi_0\rangle'$, these equations will define a different set of linear operators acting on a different subspace ${\cal H}_{\Psi'_0}$. The arguments of \cite{Almheiri:2013hfa, Marolf:2013dba} imply that it is impossible to upgrade the operators $\widetilde{\cal O}$ to fixed linear operators defined once and for all on the entire Hilbert space, with the property that they satisfy the desired equations \eqref{todefa} on typical states. However for a given microstate $\rsz$, and all of its small excitations in ${\cal H}_{\Psi_0}$, we can work with the state-dependent operators $\widetilde{\cal O}$ as defined on ${\cal H}_{\Psi_0}$ by the equations \eqref{todefa}.

We assumed that the state $\rsz$ is an equilibrium state. However the subspace ${\cal H}_{\Psi_0}$ will also contain non-equilibrium states. One class of non-equilibrium states are those of the form $U({\cal O})\rsz$, where $U$ is a unitary made
out of the elements ${\cal O}$ of the small algebra\footnote{We assume that the expression of $U$ in terms of ${\cal O}$ is such that if we expand $U$ in a power series in ${\cal O}$, the operator $U$ can be well approximated by low powers such that it effectively does not exceed the cutoff of the algebra ${\cal A}$. \label{footunit}}. Another class, which is the main emphasis of this paper, are states of the form 
\be
\label{bbbb}
\rs\equiv U(\tO)\rsz
\ee
where again $U$ is a unitary constructed out of the operators $\tO$ defined by \eqref{defto}. Notice that these operators are state-dependent and they are only defined within the subspace ${\cal H}_{\Psi_0}$.

We consider the expectation value of an operator $A \in {\cal A}$ on such a state
\begin{align}
\label{seemseq}
\ls A \rs =  \lsz U^\dagger(\tO) A U(\tO) \rsz   = \lsz A \rsz
\end{align}
In the second equality we used the large $N$ commutator $[A,\tO]=0$. Hence expectation values of elements of the small algebra ${\cal A}$ on the state $U(\tO)\rsz$ will be almost the same as those in the equilibrium state $\rsz$ and in particular they will be time-independent. This may seem to suggest that the state $U(\tO)\rsz$ is an equilibrium state. On the other hand if we insert factors of the Hamiltonian in the correlator we find
\be
\label{corrh}
\ls A H \rs \neq \lsz A H \rsz
\ee
since $H$ does not commute with $\tO$, which follows from \eqref{comK} and  \eqref{todefa}. In particular, unlike the correlator on the equilibrium state on the RHS of \eqref{corrh} which is static, the correlator on the LHS may be time dependent indicating that the state $\rs$ is not typical and is not an equilibrium state.

Now, using the definition of the operators \eqref{defto} the state $\rs \equiv U(\tO) \rsz$ can also be written as
\be
\label{bbbc}
\rs = e^{-{\beta H \over 2}} U({\cal O}^\dagger) e^{{\beta H \over 2}} \rsz
\ee
Since \eqref{bbbb} and \eqref{bbbc} are the same as states, then the relations \eqref{seemseq}, \eqref{corrh} continue to hold for \eqref{bbbc}. The important point is that when written as \eqref{bbbc} the state is created by acting with ordinary, state-independent operators. 

In the rest of the paper we will consider the states \eqref{bbbc} in their own right, independent of the fact that they were inspired by the state-dependent construction described above. The main intuition is that while states of the form $U({\cal O}) \rsz$ describe black holes with excitations in the exterior of the horizon, conjugating the unitary by the factors $e^{\pm {\beta H \over 2}}$ produces a state where the wavepacket is localized in a complementary region, which can be naturally identified with a region contained entirely behind the horizon. We proceed by describing a toy-model of this phenomenon in flat space QFT without gravity.

\section{Warmup: states behind the horizon in Rindler space}
\label{secrind}

Consider a non-gravitational relativistic 4d QFT in the Minkowski vacuum $|0\rangle$ and consider the Rindler decomposition. The right Rindler wedge $R$ is defined as the region $x^1>0, |x^1|>|t|$, while the left wedge $L$ is defined by $x^1<0, |x^1|>|t|$. 
We consider the algebras ${\cal A}_R, {\cal A}_L$ corresponding to operators localized in the two respective wedges. The two algebras commute. 

\begin{figure}[!t]
\begin{center}
\includegraphics[width=.6\textwidth]{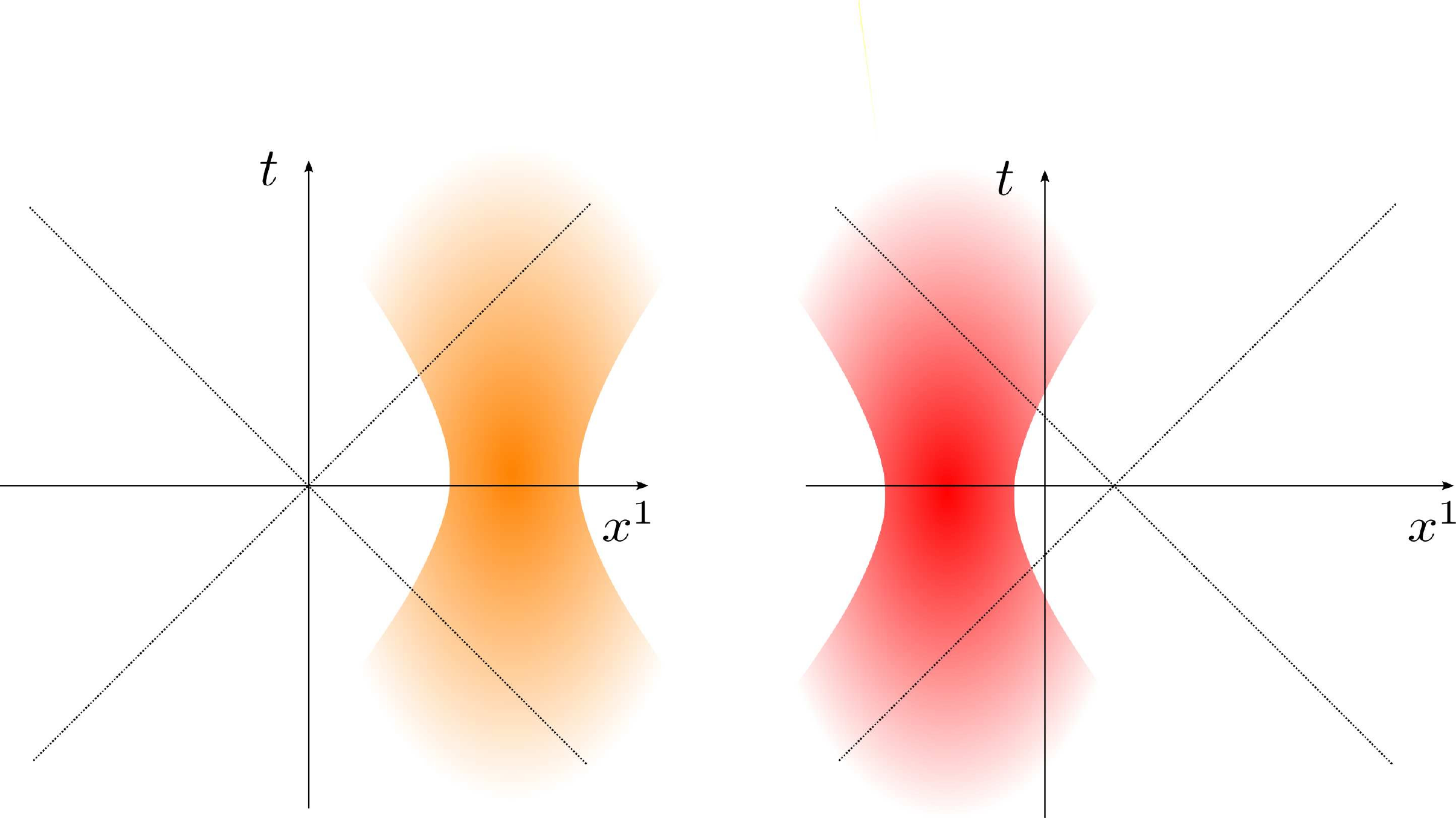}
\caption{ Left: a state of the form $U_R\vac$, where $U_R$ is unitary localized in the right wedge. Right: the corresponding state $e^{-\pi M} U_R \vac$, where $M$ is the Lorentz boost generator on the $t$-$x^1$ plane. The shaded regions show schematically the points where correlators differ from vacuum correlators. The state on the right represents an excitation which, from the point of view of an observer in $R$, remains behind the horizon for all time.}
\label{rindsketch}
\end{center}
\end{figure}

We imagine an observer living in wedge $R$ and we want to consider states with excitations behind the Rindler horizon: by this we mean that the excitation stays and evolves entirely behind the horizon and never crosses into region $R$ --- see second sub-figure in \ref{rindsketch}.

We will review how the Tomita-Takesaki construction can be applied to this case leading to the well known result from the Unruh effect, that the Lorentz boost generator $M$  along the $t$-$x^1$ directions plays the role of the modular Hamiltonian for the decomposition into ${\cal A}_R, {\cal A}_L$. We will also see that if we start with a state $U_R \vac$ localized in the right wedge, then the state $e^{-\pi M} U_R e^{\pi M}\vac$ is localized in the left wedge\footnote{As seen in figure \ref{rindsketch}, the more precise statement is
that $U_R\vac$ is localized entirely {\it outside} the left wedge, and $e^{-\pi M} U_R e^{\pi M}\vac$ is localized entirely outside the right wedge.}. Notice the analogy to \eqref{bbbc}. This result can be easily derived in free field theory using the Bogoliubov transformation between Minkowski and Rindler modes, as reviewed towards the end of this section, but the result is more general and holds also for interacting QFTs.

According to the Reeh-Schlieder theorem \cite{Reeh1961}, by acting with elements of the algebra ${\cal A}_R$ on $|0\rangle$ we can generate a dense subspace of the full Hilbert space. Moreover, it is impossible to find annihilation operators for $|0\rangle$ which are entirely localized in the right wedge. Hence $|0\rangle$ is a cyclic and separating vector for the algebra ${\cal A}_R$. 
Thus the Tomita Takesaki construction can be applied, which is essentially captured by the analysis of Bisognano and Wichmann \cite{Bisognano:1975ih,Bisognano:1976za}. They showed that if we define an antilinear operator $\cS$ such that $\cS A_R |0\rangle = A_R^\dagger |0\rangle$, for any operator $A_R$ constructed out of fields in $R$, then the operator $\cS$ can be expressed as
\be
\label{bws}
\cS = \Theta U(R_1(\pi)) e^{-\pi M}
\ee
where $\Theta$ is the CPT operator, $U(R_1(\pi))$ is a rotation  by $\pi$ around the $x^1$ axis and $M$ is the Lorentz boost generator on the $t$-$x^1$ plane. Notice the last factor is a complexified Lorentz boost. To understand how \eqref{bws} is derived, we can take $A_R$ to be a local scalar operator ${\cal O}(x)$ in the right wedge. By temporarily ignoring issues of the domains of operators, we can see that the sequence of the complexified Lorentz boost, the rotation by $\pi$ around $x^1$ and the CPT operator, map the point $x$ back to itself, with an additional conjugation of the operator, that is $\cS {\cal O}(x) \vac = {\cal O}^\dagger (x) \vac$. In \cite{Bisognano:1975ih,Bisognano:1976za} the convergence of this mapping was established and it was extended to more general operators in the right wedge.

From \eqref{bws}  we find that
$$
\Delta = \cS^\dagger \cS = e^{-2\pi M}
$$
We thus find that the modular Hamiltonian, defined as $K \equiv -\log\Delta$, is proportional to the Lorentz boost generator
$$
K = 2\pi M
$$
The coefficient $2\pi$ is the inverse temperature $T={1\over 2\pi}$ of the Unruh effect.

We also find that the modular conjugation operator \eqref{defj} is given by
\be
\label{jrind}
J = \Theta U(R_1(\pi))
\ee
which obeys $J = J^\dagger = J^{-1}$. Notice that if ${\cal O}_R$ is a local operator in the wedge $R$ then $J{\cal O}_R J$ is localized in the left wedge and more generally the results
of \cite{Bisognano:1975ih,Bisognano:1976za} imply for the algebras
\be
\label{algmap}
J {\cal A}_R J = {\cal A}_L
\ee

Now, suppose we want to consider a state with an excitation localized outside the left wedge $L$. Such a state can be constructed by acting with a unitary operator made out of operators from the algebra ${\cal A}_R$. We denote such a unitary operator by $U_R$. We think of it as being the exponential $e^{i {\cal O}_R}$ of some smeared local operator ${\cal O}_R$ in region $R$. We can also think of it as the operator that we get by modifying the Hamiltonian by turning on a localized source for a while in the region $R$. The state $U_R|0\rangle$ contains some excitations on top of the vacuum, however these excitations cannot be detected in region $L$. We can see that by computing the expectation value of any operator ${\cal O}_L\in{\cal A}_L$, on the state $\rs$. We have
$$
\ls {\cal O}_L \rs = \langle 0| U_R^\dagger {\cal O}_L U_R |0\rangle = \langle 0| {\cal O}_L |0\rangle
$$
where we used $[{\cal O}_L,U_R]=0$, since the support of the two operators is spacelike. From this we see that all correlators of elements of ${\cal A}_L$ are the same in the state $\rs$ as in $|0\rangle$.\footnote{Notice that this is generally true only for unitary operators $U_R$. For example, if we act with a general Hermitian operator $\phi_R$ localized in $R$ and consider the 
state $c \phi_R \vac$, where $c$ is a normalization factor, then despite the fact that $[{\cal O}_L,\phi_R]=0$ correlators in the left wedge on this state are different from those in $\vac$.} 

 Now we consider states of the form
\be
\label{newstatesR}
\rsp\equiv e^{-\pi M} U_R|0\rangle
\ee
Since $M|0\rangle = 0$ we can equivalently write the state as $\rsp = e^{-\pi M}U_R e^{\pi M}\vac$, to make the analogy with \eqref{introstatesa} more clear.
These states contain excitations which are not detectable in region $R$. 
More precisely we will see that all correlators of operators in region $R$  on the state $\rsp$ are the same as those in the vacuum. Notice that the operator $e^{-\pi M}$ is unbounded, as the spectrum of the Lorentz boost generator $M$ is $(-\infty, \infty)$. However the state $U_R |0\rangle$ is in the domain of the operator $e^{-\pi M}$. This follows from the analysis of \cite{Bisognano:1975ih,Bisognano:1976za} and is discussed in more detail in appendix \ref{domainboost}.

To prove that the state \eqref{newstatesR} is localized outside the wedge $R$, we insert a factor of $J^2 =1$, where $J$ is given in \eqref{jrind} to get
\be
\label{rindul}
\rsp = J^2 e^{-\pi M} U_R |0\rangle = J \cS U_R \vac = J U_R^\dagger \vac = (J U_R^\dagger J) \vac = U_L \vac
\ee
where $U_L$ is a unitary operator in the left wedge. Here we used that $J\vac=\vac$ and the property $\cS A_R \vac = A_R^\dagger \vac$, which was the defining property of $S$. To check that $U_L$ is indeed unitary, we first notice from \eqref{algmap} that $U_L\equiv J U_R^\dagger J$ is an operator localized in $L$. We also have
$U_L^\dagger U_L = (J U_R J)^\dagger (J U_R^\dagger J) = 1$, where we used $J^\dagger = J= J^{-1}$ and that $U_R$ is unitary. Hence $U_L$ is indeed a unitary in the left wedge.

Given that the state \eqref{newstatesR} can also be written as $\rsp = U_L \vac$, where $U_L$ is unitary and $[U_L,{\cal O}_R]=0$,
the desired property is obvious
\be
\label{ttkms}
\ls {\cal O}_R \rs = \langle 0| U_L^\dagger {\cal O}_R U_L |0\rangle = \langle 0| {\cal O}_R \vac
\ee
for any operator ${\cal O}_R$ in the right wedge, which may even be non-local or  the product of many local operators in wedge $R$.

This shows that we can prepare states which have excitations only outside region $R$, by acting with unitaries made out of operators in $R$ and then further multiplying by the factor $e^{-\pi M}$. Of course $M$ has support both on $L$ and $R$. 
Notice that while in this way we can construct the states with excitations in region $L$, we cannot construct the operators acting on $L$. This can be understood by noticing that while $e^{-\pi M}U_R$ (or $e^{-\pi M } U_R e^{\pi M}$) creates
excitations purely outside region $R$ when acting on the vacuum $|0\rangle$, this will no longer be true if we act with it on a more general excited state, since the proof presented above will not go through for general states\footnote{For example, if a state contains excitations in the left wedge, then it may not even be in the domain of definition of the operator $e^{-\pi M}$, as we discuss below.}. Equivalently, $e^{-\pi M}U_R e^{\pi M}$ does not commute as an operator with general operators ${\cal O}_R \in {\cal A}_R$, so it is not an element of ${\cal A}_L$.

As discussed in the previous section something similar is happening in the case of a thermal quantum system, which may be dual to a black hole in AdS. The role of the algebra ${\cal A}_R$ is played by the algebra of simple operators, while the role of the modular Hamiltonian $K$ is played by the physical Hamiltonian generating time translations. Both for the AdS black hole and for Rindler space the states with excitations living entirely on the other side of the horizon are constructed as
$$
e^{-{K \over 2}} U({\cal O}) e^{K \over 2}\rs
$$
where $K$ is the modular Hamiltonian. We could just write this as $e^{-{K \over 2}} U({\cal O})\rs$ since $K$ annihilates $\rs$.

The analogy is not complete: in the case of the black hole we are applying the Tomita Takesaki construction in an approximate sense, since the small algebra ${\cal A}$ that we defined is not a proper algebra, unlike ${\cal A}_R$ in the case of QFT in flat space\footnote{Even in QFT in flat space it is nontrivial to go from the polynomial algebra of smeared operators, which are generally unbounded, to a von Neumann algebra of localized, bounded operators.}. As mentioned before, this is important as a realization of the idea of black hole complementarity. 

Another comment is that in the case of non-gravitational Rindler space the modular Hamiltonian $2\pi M$ can be constructed only if we have access to both wedges. In the case of a 1-sided black hole the modular Hamiltonian is determined by the CFT Hamiltonian, which {\it naively} seems to be related to the exterior region near the boundary, however due to the gravitational dressing it effectively has support also in the interior.

Yet another difference is that in non-gravitational Rindler space QFT we have only one state, the Minkowski vacuum $\vac$, corresponding to a system in ``equilibrium'', while in the case of the thermal system any typical  pure state $\rsz$, dual to a specific black hole microstate, can play the role of the reference equilibrium state. 
\vskip10pt
\noindent{\bf Free field theory}
\vskip10pt
Consider a free field in the Rindler decomposition. The field is expanded in right and left Rindler oscillators which we denote respectively  as $a_\omega, \ta_\omega$. Here we write only the index of the Rindler frequency $\omega$ and suppress the index of the transverse momentum. These oscillators obey the standard algebra
$$
[a_\omega, a_{\omega'}^\dagger] = \delta(\omega-\omega') \quad,\quad [\ta_\omega, \ta_{\omega'}^\dagger] = \delta(\omega-\omega')
$$
with other commutators equal to zero. We also have the Unruh thermal occupation expectation values with $\beta = 2\pi$ leading to
\be
\label{runru}
\langle 0|a_\omega^\dagger a_{\omega'}|0 \rangle = \langle 0|\ta_\omega^\dagger \ta_{\omega'}|0 \rangle = \delta(\omega-\omega') {e^{-2\pi \omega} \over 1-e^{-2\pi \omega}}
\ee
In terms of these oscillators the Lorentz generator has the form
$$
M = \int_0^\infty d\omega \, \omega \left(a_\omega^\dagger a_{\omega}- \ta_\omega^\dagger \ta_{\omega} \right)
$$
As an example, start with a unitary on the right wedge expanded to linear order in $\theta$
$$
U_R = e^{i \theta (a_\omega + a_\omega^\dagger)}\vac = \vac + i \theta (a_\omega + a_\omega^\dagger) + O(\theta^2)
$$
To be more precise we should have smeared the modes, but the logic would be the same. We have
$$
e^{-\pi M} U_R \vac =  \vac + i \theta (e^{\pi \omega} a_\omega + e^{-\pi \omega}a_\omega^\dagger)\vac + O(\theta^2)
$$
Using the Bogoliubov relations
$$
(\ta_\omega -e^{-\pi \omega} a^\dagger_\omega) \vac =0
$$
$$
(\ta_\omega^\dagger- e^{\pi \omega} a_\omega)\vac = 0
$$
we can write the state above as
$$
e^{-\pi M} U_R \vac =  \vac + i \theta (\ta_\omega^\dagger + \ta_\omega^\dagger)\vac + O(\theta^2)
$$
which is the the same as the expansion to linear order in $\theta$ of a unitary localized in the left region
$$
U_L \vac = e^{i \theta (\ta_\omega^\dagger + \ta_\omega^\dagger)} \vac
$$
More generally, according to \eqref{rindul}, we have that
$$
e^{-\pi M} U_R \vac = U_L \vac
$$
\vskip10pt
\noindent{\bf The eternal AdS black hole}
\vskip10pt
Another example where a similar algebraic structure appears is the eternal AdS black hole, dual to two non-interacting CFTs in the thermofield (TFD) entangled state
$$
|{\rm TFD}\rangle = \sum_E {e^{-\beta E/2}\over \sqrt{Z}}|E,E\rangle
$$
If in this case we take the algebra ${\cal A}$ to be the algebra of operators in the right CFT, then we find that 
$$
\Delta = e^{-\beta (H_R-H_L)}\quad, \quad K = \beta (H_R-H_L)
$$
and the antilinear modular conjugation operator $J$ interchanges operators between the two CFTs 
$$
J {\cal O}_R J = {\cal O}_L
$$

We can create excitations which are localized in the right AdS region as
$$
U({\cal O}_R) |{\rm TFD}\rangle 
$$
If we conjugate the operator using $\Delta^{1/2} = e^{-{K \over 2}}$, and following similar steps as the one we used for Rindler space,  we have the relation 
\be
\label{modconjt}
e^{-{\beta (H_R-H_L) \over 2}}U({\cal O}_R)  |{\rm TFD}\rangle  = U({\cal O}_L) |{\rm TFD}\rangle
\ee   
so the state can also be written as a unitary made out of the left CFT operators, which means that the excitations is completely localized in the left CFT. 

\section{Equilibrium states and spontaneous fluctuations}
\label{eqmaintext}

Returning to statistical mechanics, in this section we review some useful concepts. A more complete discussion is presented in appendix \ref{statmech}.

We want to define the notion of an equilibrium {\it pure} state for an isolated quantum system. We assume that the system is bounded and has  discrete spectrum. Before considering pure states we mention two commonly used equilibrium {\it mixed} states, the canonical one $\rho_\beta = {e^{-\beta H} \over Z}$ and the microcanonical $\rho_m = {{\mathbb P}_{E_0} \over {\cal N}}$.  Here ${\mathbb P}_{E_0}$ is a projector on the subspace spanned by energy eigenstates which lie within a small window centered around $E_0$ and ${\cal N} = e^{S}$ is the dimensionality of this subspace. Since $[\rho_\beta,H]=[\rho_m,H]=0$, all observables are exactly time-independent on both of these mixed states. Notice that expectation values in these two mixed states  differ by powers of $1/S$
\be
\label{microvscanf}
\Tr[\rho_m A] = \Tr[\rho_\beta A] + O(1/S)
\ee
Here we assume $||A|| \sim O(S^0)$. Here $\beta$ and $E_0$ are related by ${\partial S \over \partial E}|_{E_0} = \beta$.

To define the notion of equilibrium for a pure state we require two conditions: i) that correlators of simple observables on the state are approximately time-independent and ii) that their values are close to the thermal correlators on $\rho_\beta$, or equivalently up to $1/S$ corrections, to microcanonical correlators on $\rho_m$.

Equilibrium pure states are related to {\it typical} pure states. A typical state is defined by considering a window of energies centered around $E_0$ and writing down a superposition of energy eigenstates in that window
\be
\label{micropuref}
|\Psi\rangle = \sum_i c_i |E_i\rangle
\ee
The coefficients $c_i$ are selected randomly, subject to the normalizability condition $\sum_i |c_i|^2 =1$. This defines the notion of a typical pure state corresponding to the micrononical ensemble, since in \eqref{micropuref} we have included only eigenstates $|E_i\rangle$ from a particular energy window.  It is easy to show that for such typical pure states we have

\be
\label{purevsmixedf}
\ls A \rs = \Tr[\rho_m A] + O(e^{-S/2})
\ee
where again we assume that the norm $||A||$ scales like $S^0$. See appendix \ref{statmech} for a more precise definition of typical states and for the precise interpretation of \eqref{purevsmixedf}.

For the following discussion it is useful to consider the expectation value and the energy spread of a pure state
$$
E_0 \equiv \ls H \rs
$$
\be
\label{enspreadf}
(\Delta E)^2 \equiv  \ls H^2 \rs - \ls H \rs^2
\ee

We  assume that the observables $A$ that we will consider obey the Eigenstate Thermalization Hypothesis (ETH) \cite{peresETH, deutsch, srednicki1999approach}, which postulates that the matrix
elements of observables on energy eigenstates have the following structure
\be
\label{ethdeff}
\langle E_i| A |E_j\rangle = f(E_i) \delta_{ij} + R_{ij} e^{-S/2} g(E_i,E_j)
\ee
where the functions $f,g$ are smooth functions of the energy and of magnitude $O(S^0)$, while $R_{ij}$ are complex coefficients with magnitude of $O(1)$ and erratic phases. Here $S$ is the entropy at the relevant energy\footnote{Which can be either $E_i$ or $E_j$, by absorbing an appropriate factor in $g$.}. 

We can now see that typical states are equilibrium states. We consider the time dependence of any observable on a typical state. Using \eqref{ethdeff} we find
\be
\label{ethexpf}
\langle A(t) \rangle = \sum_i |c_i|^2 f(E_i)  + \sum_{i\neq j} e^{-i (E_j-E_i)t} c_i^* c_j R_{ij}e^{-S/2} g(E_i,E_j) 
\ee
For typical states the second term is of order $O(e^{-S/2})$ because of cancellations of random phases. To understand the scaling, the coefficients $c_i$ scale like $e^{-S/2}$, we have an explicit factor of $e^{-S/2}$ from the size of the matrix elements of $A$ and
we get another factor of $\sqrt{e^{2S}}$ since we are summing over $e^{2S}$ random phases. In this argument it was important that for typical states, the phases of the coefficients $c_i$ are uncorrelated to the erratic phases of $R_{ij}$.

Hence we find that the expectation value of $A$ is almost time independent. Moreover this time independent value is given by the first term in \eqref{ethexpf}. According to the ETH the coefficient $f(E_i)$ is a smooth function of the energy. So if the state has highly peaked energy the time independent value will be almost equal to the expectation value in the microcanonical ensemble, and up to $1/S$ corrections equal to the thermal expectation value at the appropriate temperature.

We emphasize that for pure states the notion of equilibrium may depend on the time range that we probe the state. If we start with a state that looks like an equilibrium state and  if we wait long enough, there may be a time $t_0$ where the phase factors $e^{-i(E_j-E_i)t} c_i^* c_j $ in \eqref{ethexpf} get correlated with $R_{ij}$ in such a way that the off-diagonal terms become as important as the diagonal term, for a certain time interval. When this happens the state which used to look like an equilibrium state at  early times, will temporarily get out of equilibrium. This can lead to nontrivial time-dependence of observables around the time $t=t_0$. Upon further time evolution the phases will decohere and the state will start to look again like an equilibrium state. If we look at the entire history of the state, this has the interpretation as a state which undergoes a {\it spontaneous non-equilibrium fluctuation} around $t=t_0$. We notice that whether this will happen or not throughout the history of the state, depends on the coefficients $c_i$. 

We also stress that such spontaneous fluctuations can only happen when the spread $\Delta E$  of \eqref{enspreadf} is appreciably large. This follows from the basic inequality
\be
\label{dedtf}
{1\over 2}\left|{d \langle A\rangle  \over dt} \right|\leq \Delta A\, \Delta E
\ee in quantum mechanics, where $\Delta A \equiv \sqrt{\ls A^2\rs -
\ls A \rs^2}$. This inequality implies that if the spread $\Delta E$ of a pure state is too small then there will not be interesting spontaneous fluctuations in the history of the state even if we wait exponentially long. 

\section{Excitations of equilibrium states and the black hole geometry}
\label{mainsection}

In this section we consider three classes of states and their bulk interpretation in AdS/CFT: i) equilibrium states ii) states with excitations visible outside the horizon iii) states with excitations hidden behind the horizon.

\subsection{Equilibrium states and static black hole geometry}
\label{standeq}

We take a  pure state $\rsz$ selected from the microcanonical ensemble centered around energy $E_0$ and with spread $\Delta E$. If the width $\Delta E$ is extremely small, then the state will look almost like an energy eigenstate, and given \eqref{dedtf}, reasonable
observables will not have any interesting time variation, even if we wait 
for exponentially long times. If the spread $\Delta E$ is sufficiently large, 
then the state may have time-dependence and undergo spontaneous fluctuations away
from equilibrium. However, as reviewed in appendix \ref{eqreview} most states and for most of their 
history will look like equilibrium states even if $\Delta E$ is appreciable.

For such  typical pure states correlators of operators of the algebra ${\cal A}$  
are time-independent up to exponentially small corrections
\begin{equation}
\label{eqconda}
{d\over dt} \lsz A(t) \rsz = O(e^{-S/2})
\end{equation}
as we already explained after \eqref{ethexpf}. The same conclusion can be reached for correlators involving operators from the algebra ${\cal A}$ including the Hamiltonian $H$, i.e.
\begin{equation}
\label{eqcondbb}
{d\over dt} \lsz A(t) H\rsz = O(e^{-S/2})
\end{equation}
In fact we can derive \eqref{eqconda}, \eqref{eqcondbb}  even without invoking the ETH. We reviewed in \ref{purevsmixedf} that correlators on typical pure states from the microcanonical distribution are  exponentially close to correlators on the microcanonical mixed state $\rho_m = {P_{E_0} \over {\cal N}}$. In the microcanonical mixed state $\rho_m$ the time derivative of the expectation value of any operator is exactly equal to zero, since $[\rho_m, H]=0$. From these two arguments the conclusions \eqref{eqconda}, \eqref{eqcondbb}  follow. 

Notice that here we are talking about the time-dependence of the expectation value of the operator $A(t) = e^{i H t} A e^{-i H t}$, where $A$ is a fixed operator in the Schroedinger picture. We could for instance consider an $n$-point function by taking $A =
e^{i H t_1} {\cal O} e^{-i H t_1}\ldots e^{i H t_n} {\cal O} e^{-i H t_n}$, where we think of $t_i$'s as fixed parameters defining the operator $A$, and consider the $t$-dependence of
\be
 \lsz {\cal O}(t+t_1)\ldots {\cal O}(t+t_n) \rsz
\ee
The statement that $\rsz$ is in equilibrium means that the $n$-point function does not depend on the overall time $t$, but of course it can depend nontrivially on the time differences $t_i$. 

To summarize, correlators of ${\cal A}$  on typical pure states from the microcanonical ensemble are exponentially close to correlators on the microcanonical mixed state $\rho_m$ and appear to be in equilibrium. In addition we can approximate the correlators in $\rho_m$ by those in the canonical mixed state $\rho_\beta = Z^{-1}e^{-\beta H}$, up to $1/S$ corrections
\be
\label{canmicro}
\Tr[{\rho_m} A ] = {\Tr[e^{-\beta H} A] \over Z}+ O(1/S)
\ee
The thermal correlators on the RHS can be computed by analytic continuation starting from the Euclidean theory with the time compactified on a circle of length $\beta$.

In a large $N$ CFT with a gravitational dual, these considerations allow us to determine the leading order correlators in $1/N$, on a typical pure state in the high temperature deconfined phase. We start by the basic prediction  by AdS/CFT, that Euclidean thermal correlators can be computed by a perturbative expansion in Witten diagrams around the Euclidean black hole. By analytic continuation we get the real-time thermal correlators on the mixed state $\rho_\beta= Z^{-1}e^{-\beta H}$. We assume that upon analytic continuation to real time, and provided that 
the time separation between the local operators composing $A$ is not too large, the $1/N$ corrections of the Euclidean correlator remain subleading. From the real time correlators on $\rho_\beta$ we can derive the leading order real time correlators on the microcanonical ensemble $\rho_m$. Finally from \eqref{purevsmixedf} we can also get the leading order correlators on a typical pure state. Then we find that real-time
correlators of single trace operators on a typical equilibrium pure state factorize to 2-point functions\footnote{Here we assume that we have defined ${\cal O}$ so that its 1-point function is zero.}
\begin{align}
\label{puretherm}
\lsz {\cal O}(t_1,x_1)\ldots  {\cal O}(t_{2n},x_{2n}) \rsz = & \lsz {\cal O}(t_1,x_1) {\cal O} (t_2,x_2)\rsz\ldots \lsz {\cal O}(t_{2n-1},x_{2n-1}) {\cal O} (t_{2n},x_{2n})\rsz\cr
& + {\rm permutations} + O(1/N)
\end{align}
where the 2-point functions  are almost equal to the thermal ones
\be
\label{purethermb}
\lsz {\cal O}(t_1,x_1) {\cal O} (t_2,x_2)\rsz = {\Tr}\left[e^{-\beta H} {\cal O}(t_1,x_1) {\cal O}(t_2,x_2)\right] + O(1/N)
\ee
The approximation of large $N$ factorization is reliable only when the time differences are not too large, or too small. For example, in out of time order correlators for which $\Delta t$ approaches the scrambling time $\beta \log S$, the subleading $1/N$ corrections may become as large as the leading order terms \cite{Shenker:2013pqa,Maldacena:2015waa}. Also, $1/N$ corrections may get enhanced in Lorentzian correlators in particular kinematic regimes related to lightcones. 

Equations \eqref{puretherm},\eqref{purethermb} imply that, to leading order in $1/N$, the dual geometry of a typical pure state $\rsz$ is determined by the analytic continuation of the Euclidean black hole.
In particular it is clear that the dual spacetime will contain at least the region  outside a static black hole horizon, with the quantum fields on this spacetime placed in the usual Hartle-Hawking thermal state. This can be made precise by considering the HKLL construction \cite{Hamilton:2006az} of local operators in the bulk and showing that their correlators agree with those computed from semiclassical gravity \cite{Papadodimas:2012aq}. Notice that from the ETH, the same conclusion applies even if  $\rsz$ is an exact energy eigenstate. 

The question of how to extend the spacetime behind the past and future horizons has been subject of discussion, with proposals \cite{Almheiri:2013hfa, Marolf:2013dba} that the diagram may have to be terminated on the horizons with a firewall or some kind of fuzzball. In  \cite{Papadodimas:2012aq,Papadodimas:2013b,Papadodimas:2013,Papadodimas:2013kwa,Papadodimas:2015xma,Papadodimas:2015jra}  a proposal was given of how the CFT may describe the extended Penrose diagram of the static black hole.  We will argue that the existence of certain non-equilibrium states in the CFT, which can be identified with excitations moving in a region {\it different from the exterior of the black hole}, provides support to the idea that we should extend the spacetime diagram to the interior region, and part of what would be the left asymptotic region, as depicted in figure \ref{fig1}.

Before we move to non-equilibrium states, we mention that computing the subleading corrections in $1/N$ on a typical pure state is more complicated. The computation of $1/N$ corrections to thermal Euclidean correlators can be done by computing interacting Witten diagrams on the Euclidean black hole. These can be analytically continued to real time, thus allowing us to compute $1/N$ corrections to real-time correlators on the canonical mixed state $\rho_\beta $. We would like to relate those to the subleading corrections in $\rho_m$ and eventually use the fact that correlators on $\rho_m$ and on a typical pure state are exponentially close \eqref{purevsmixedf}. However, correlators in the two ensembles $\rho_m, \rho_\beta$ differ by $1/N$ terms \eqref{canmicro}\footnote{For big AdS black holes dual to the deconfined phase of the ${\cal N}=4$ SYM in the 't Hooft limit we have $S \sim c N^2 T^4R^4$.}, which appear at the same order with the $1/N$ corrections coming from interactions in the bulk. It might be possible to disentangle the two by an inverse Laplace transform from one ensemble to the other, but this has not yet been analyzed in detail in relevant examples.

\subsection{Standard non-equilibrium states --- excitations outside the horizon}

We now want to consider a standard class of non-equilibrium states, corresponding to a black hole with a few excitations outside the horizon (on top of the equilibrium Hawking radiation surrounding a big AdS black hole).
We start with an equilibrium state $|\Psi_0\rangle$
and we act on it with a unitary operator of the form $U({\cal O})$, which is made out of operators of the algebra ${\cal A}$ of single trace operators localized around a given time $t_0$. In AdS/CFT we may be interested in exciting the state by an HKLL operator  \cite{Hamilton:2006az}, corresponding to creating a wavepacket in the bulk, in the region outside the horizon. For simplicity we assume that the energy change due to $U({\cal O})$ scales like $N^0$, so that the backreaction on the classical geometry is not significant.

We consider the state
\be
\label{standeq}
\rs = U({\cal O}) |\Psi_0\rangle
\ee
This generally behaves like a non-equilibrium state around the time $t=t_0$ and if we compute correlators on this state near $t_0$ we will have
$$
{d\over dt}\ls A(t) \rs \neq 0
$$
For example, if we consider a perturbation of the form $U=e^{i \theta {\cal O}(t_0)}$, where $\co$
is Hermitian, we then find
\be
\label{noneqch}
\ls {\cal O}(t) \rs =  \lsz {\cal O}(t) \rsz  + i\theta \lsz [{\cal O}(t), {\cal O}(t_0)] \rsz + O (\theta^2)
\ee
While the first term is almost time-independent \eqref{eqconda}, the second term will generally be non-zero and time-dependent. Hence by measuring the time-dependence of the expectation value of ${\cal O}(t)$ on this state we can see that it is a non-equilibrium state. However, at  very late, or very early times, the state will equilibrate again: the linear term in $\theta$  in equation \eqref{noneqch} is proportional to the imaginary part of the 2-point function $\lsz {\cal O}(t) {\cal O}(t_0) \rsz$, which decays
exponentially in the time-difference $|t-t_0|$. This is a more general conclusion, which we expect to hold not only to linear order in $\theta$: under time evolution an atypical, i.e. non-equilbrium state, will eventually start to look like a typical state and will equilibrate.

Notice that not all unitaries made out of ${\cal O}$ turn equilibrium into non-equilibrium states \cite{Almheiri:2013hfa,Harlow:2014yoa}. If a unitary $U$ commutes with the Hamiltonian $H$, then the state $U({\cal O})\rsz$ will be as typical as $\rsz$ and time correlators on it will be time-independent. Hence it will remain an equilibrium state. Various aspects of the bulk interpretation of such states have been discussed in \cite{Almheiri:2013hfa,Harlow:2014yoa,Papadodimas:2015jra,Marolf:2015dia,Raju:2016vsu,Raju:2017ost}. From now on we will consider
only $U$'s which lead to non-equilibrium states.
 
In order to fully characterize the history of the state \eqref{standeq}, including both the past and the future, we have to clarify the interpretation of \eqref{standeq}. There are two possibilities: 
 
a) As an ''actively perturbed state``: at time $t=t_0$ we quench the system by modifying the Hamiltonian of the theory by adding a source term for a very short period, schematically
$$
H(t) = H_0 +  j(t) {\cal O}(t)
$$
where the source $j(t)$ is highly peaked around $t=t_0$. Ignoring the short time that it takes for the perturbation to act, then for $t<t_0$ the state of the system is $\rsz$, while for $t>t_0$ the state is to leading order $\rs=U({\cal O})\rsz$. In this case \eqref{noneqch} is only valid for $t>t_0$. The AdS/CFT interpretation of the actively perturbed states is that we perturb the CFT, which used to be in the state $\rsz$ for $t<t_0$, by some source near $t=t_0$, injecting a wavepacket or shockwave towards the black hole.

b) As an ''autonomous state``: in this case we  do not actively perturb the system at $t=t_0$. Instead we simply consider the state \eqref{standeq} as a state in the Heisenberg picture with respect to the original unperturbed Hamiltonian $H_0$.

These two interpretations agree on what the state of the system is  for $t>t_0$, but clearly they differ for $t<t_0$. In either of the two interpretations the state right after $t_0$ is given by \eqref{standeq}.
Seen as an actively perturbed state, the full interpretation is that we start with a system in equilibrium, we excite it at $t=t_0$ and then it equilibrates again at late times. Seen as an autonomous state, we notice that not only at very late, but also at very  early times, the state looks like an equilibrium state, while at $t\approx t_0$ the state exhibits nontrivial time-dependence.
Hence the state \eqref{standeq} is one which is fine-tuned to undergo a spontaneous fluctuation at time $t=t_0$.

We emphasize   that this spontaneous fluctuation is not the same thing as the thermal fluctuations of the Hawking atmosphere around the black hole. The latter are captured by the fact that the modes outside the horizon are thermally populated, which is encoded
in the moments of the thermal occupation level $Z^{-1} \Tr[e^{-\beta H} N_{\omega}^k]$, where $N_\omega = {\cal O}_\omega^\dagger {\cal O}_\omega$. On the non-equilibrium states \eqref{standeq} the expectation values of these moments are changed from those in the thermal ensemble. So on states like \eqref{standeq} quantum observables have different statistical probabilities that in the thermal ensemble.

 Seen as an autonomous state \eqref{standeq} describes an excitation which is ejected from the past horizon into the exterior AdS spacetime, reaches a maximum radial distance and falls back into the future horizon. The details of this excitation depend on the choice of the unitary operator $U({\cal O})$. Using the HKLL  \cite{Hamilton:2006az} dictionary we can find a mapping between bulk wavepackets and boundary operators creating the corresponding excitations. Extrapolating the history of these wavepackets beyond the future and past horizons, we see that they interpolate between the past and future singularities. Directly verifying the bulk interpretation  of the non-equilibrium state \eqref{standeq} behind future and past horizons, requires the construction of local operators behind the horizon, for instance using the state-dependent operators $\tO$ in the construction
 of  \cite{Papadodimas:2012aq,Papadodimas:2013b,Papadodimas:2013,Papadodimas:2013kwa,Papadodimas:2015xma,Papadodimas:2015jra}.

We can also consider states which contain more than one wavepackets, which can be written in the form $U_1({\cal O}) \ldots U_n{(\cal O})\rsz$, with $n\ll N$.  Their bulk interpretation corresponds to several wavepackets moving in the exterior of the black hole. If the unitaries are supported on spacelike separated regions, then they commute and the bulk interpretation is clear. Otherwise, we have to take the commutators into account. If the temporal separation of these unitaries is not larger than scrambling time and each of the unitaries caries energy of order $O(N^0)$, then the large $N$ expansion is reliable. By considering the possible ways of exciting the state with various unitaries we represent the space-time description of the black hole exterior.

Incidentally we mention an extreme situation of considering the state $U({\cal O})|0\rangle$, where $|0\rangle$ is the ground state of the CFT and now $U({\cal O})$ is a unitary which injects an energy of order $O(N^2)$ at time $t=t_0$. If we consider the state for $t>t_0$ it will look like a collapsing shell of matter, which may for example correspond to an AdS-Vaidya-like solution. If we think of it as an autonomous state, then for times $t<t_0$ the interpretation is that of an exploding white hole, a time-reversed Vaidya solution where a black hole emits at once all of its energy into a single ''Hawking shockwave``. At very early and very late times the bulk looks like a black hole, while at $t=t_0$ we have a shell of energy $O(N^2)$ expanding to some maximum radius in AdS and then collapsing again. Since we focus on perturbations $U({\cal O})$ which do not backreact on the classical geometry we will not consider such states.

While autonomous states like \eqref{standeq} require fine-tuned initial conditions, their existence in the Hilbert space of the CFT is clear. In principle we can actually prepare these states, by acting with the (very complicated) ''precursor'' of the operator $U({\cal O}(t_0))$ at an earlier time $t'\ll t_0$. As time evolves from $t'$ towards $t_0$ the state will look like the autonomous state $U({\cal O} (t_0))\rsz$, and the system will spontaneously 
move out of equilibrium. Alternatively, we can quench the system at $t=t_0$ by the simple operator $U({\cal O}(t_0))$  and then  wait for a Poincare recurrence time for the state to undergo the same fluctuation, now seen as a spontaneous fluctuation.

It is interesting to consider the sensitivity to the initial conditions. The state $U({\cal O}(t_0))\rsz$ is fine-tuned to undergo a spontaneous fluctuation at $t=t_0$, which means that expectation values around $t=t_0$ deviate from the equilibrium values. Suppose that we add an additional small perturbation by considering the state
$ U({\cal O}'(t_1)) U({\cal O}(t_0)) \rsz$, where $t_1\ll t_0$. If $|t_1-t_0|\gg \beta \log S$ then due to the chaotic behavior of the system the expectation value of ${\cal O}(t_0)$ will return to its equilibrium expectation value. From the bulk point of view the first particle/shockwave at $t=t_1$ pushes the second one behind the horizon long before the time $t=t_0$ \cite{Shenker:2013yza}.

Finally, it is natural to ask what is the probability for a state to undergo such a spontaneous fluctuation at some point in its lifetime. The question is, given a pure state $\rs$ with energy spread $\Delta E$, how likely is it that at some point in its lifetime it will look like $U({\cal O}) \rsz$ for some unitary made out of the algebra ${\cal A}$ and some equilibrium state $\rsz$. This would give us the statistics of the particles that the white-hole singularity will emit throughout the lifetime of the state. We do not address this question here.

\subsection{Non-equilibrium states and excitations behind the black hole horizon}

We now consider the states of main interest in this paper, which are of the form
\be
\label{newstates}
|\Psi\rangle = e^{-{\beta H \over 2}} U({\cal O}) e^{{\beta H \over 2}} |\Psi_0\rangle
\ee
where $|\Psi_0\rangle$ is a typical equilibrium state with average energy $E_0$ and spread in energy $\Delta E$, which we will assume to be small enough so that the state $|\Psi_0\rangle$ can be interpreted as a state from the  microcanonical ensemble\footnote{We also assume that $\rsz$ has compact support on the energy spectrum, so that it lies in the domain of the unbounded operator $e^{{\beta H \over 2}}$.}.  We assume that the unitary $U({\cal O})$ does not commute with $H$, so that the state $U({\cal O})\rsz$ would be a non-equilibrium state. 

We can also write \eqref{newstates}
as 
$e^{-{K \over 2}} U({\cal O})e^{{K \over 2}}\rs$ where $K=\beta(H-E_0)$ is the large $N$ 	modular Hamiltonian, see \eqref{modeq}. If the energy of the state $\rsz$ is very highly peaked around $E_0$ it can also be approximately written as $\rs = e^{-{\beta (H-E_0) \over 2}} U({\cal O}) \rsz$. Notice the analogy with the states \eqref{newstatesR} in Rindler space.

We will see that a state  of the form \eqref{newstates} is a  non-equilibrium state, even though it seems to be in equilibrium when probed by the algebra ${\cal A}$. In AdS/CFT, if we take the state $|\Psi_0\rangle$ to be a typical equilibrium black hole microstate, then we propose that the state \eqref{newstates} has the interpretation as a state with excitations localized entirely behind the horizon. This follows from the fact, shown in \eqref{proofeq}, that large $N$ correlators of single trace operators on the state \eqref{newstates} are the same as those in an equilibrium state. Hence we  have to associate to this state, the exterior geometry of a static AdS black hole with the fields in the exterior region in a thermal
density matrix, without any additional excitations. On the other hand, we will see in \eqref{noneqfi} that the state  \eqref{newstates} is out of equilibrium, so it is natural to represent these non-equilibrium excitations in the bulk as being localized behind the horizon.

To proceed, let us first introduce the shorthand notation 
\be
\label{defv}
V \equiv  e^{-{\beta H \over 2}} U({\cal O}) e^{{\beta H \over 2}} 
\ee
A first observation is that the operator $V$  is not unitary. Nevertheless, the state \eqref{newstates} is unit-normalized up to $1/S$ corrections. We see this from
\begin{align}
\label{proofnorm}
&\ls \Psi\rangle = \lsz  V^\dagger V\rsz = Z^{-1} \Tr[e^{-\beta H} V^\dagger V] + O(S^{-1})\cr
& =Z^{-1} {\rm Tr}[e^{-\beta H} e^{\beta H\over 2}U({\cal O})^\dagger e^{-{\beta H \over 2}}  e^{-{\beta H \over 2}} U({\cal O}) e^{{\beta H\over 2}}] + O(S^{-1})\cr
& = 1 +O(S^{-1}).
\end{align}
Here we used the approximation of correlators  with thermal correlators \eqref{thermalizationa} and the cyclicity of the trace. Hence while $V$ is not unitary, it gives almost unit-normalized states when acting on {\it typical, equilibrium} states\footnote{Notice that if we act with $V$ on a general non-equilibrium state, then $V$ does not preserve the norm --- not even approximately at large $S$. The proof \eqref{proofnorm} relied on the fact that $\rsz$ was an equilibrium state.}.

To make
the state \eqref{newstates} exactly unit normalized we can consider
\be
\label{normcorr}
\rs = c\, e^{-{\beta H \over 2} } U({\cal O}) e^{{\beta H \over 2}} \rsz
\ee
where the constant $c$ is selected so that $\ls \Psi\rangle =1$, i.e. 
\be
\label{normconbb}
|c|^2 = (\lsz V^\dagger V \rsz)^{-1}
\ee
Then \eqref{proofnorm} implies that at large $S$ we have $ |c|^2 = 1 + {a \over S} + ...$, where the constant $a$ is $O(1)$ and determined by the error term in \eqref{normcorr}. 
Since $|c|$ is very close to 1, sometimes we do not explicitly write this normalization factor, but it should be understood as being there in order to work with unit-normalized states.

Because $V=e^{-{\beta H \over 2}} U({\cal O})e^{{\beta H \over 2}} $ in not unitary, it is not straightforward to think of the states \eqref{newstates} as being created by actively perturbing the system by a quench. We will make some additional comments about this point in subsection
\ref{seccomments}, but for now we will continue to think of the states \eqref{newstates} as {\it autonomous} states in the Hilbert space of the theory evolving always with the unperturbed Hamiltonian $H$. 

We proceed to argue that the states of the form \eqref{newstates} have the properties that: i) they appear to be in equilibrium with respect to the algebra ${\cal A}$, even though,  ii) they are genuinely time-dependent, non-equilibrium states. These observations are motivated by the results of \cite{Papadodimas:2013}.

\vskip10pt

\noindent {\it i) State seems to be in equilibrium when probed by ${\cal A}$}
\vskip10pt
We consider an operator $A$ in the algebra ${\cal A}$ and follow the same steps as in \eqref{proofnorm}. We have
\begin{align}
\label{proofeq}
&\ls A \rs = |c|^2 \lsz V^\dagger A V \rsz = |c|^2 Z^{-1} \Tr[e^{-\beta H} V^\dagger AV] + O(S^{-1}) \cr
& =Z^{-1} {\rm Tr}[e^{-\beta H} e^{\beta H\over 2}U({\cal O})^\dagger e^{-{\beta H \over 2}} A e^{-{\beta H \over 2}} U({\cal O}) e^{{\beta H\over 2}}] + O(S^{-1})\cr
& = Z^{-1} {\rm Tr}[e^{-\beta H} A] + O(S^{-1})
\end{align}
where we used that $|c|^2 = 1 + O(1/S)$. The result \eqref{proofeq} means that the algebra ${\cal A}$ probes the state $\rs$ as if it were an equilibrium state, up to $1/S$ corrections.
In particular this means that correlators of ${\cal A}$ on this state will be time independent
\be
\label{bla4}
{d \over dt} \ls A(t) \rs = O(S^{-1})
\ee
We notice that if we try to define similar states as \eqref{newstates}, with different values of the exponents, for example $e^{-s_1 H} U({\cal O}) e^{s_2 H}|\Psi_0\rangle$ then 
these states will generally give time-dependent correlators
for the algebra ${\cal A}$, unless $s_1 = s_2 = {\beta \over 2}$.
\vskip10pt
\noindent {\it ii)  State is out of equilibrium when probed by ${\cal A}$ and  $H$ }
\vskip10pt
At this point one might think that the state $|\Psi\rangle$ defined by \eqref{newstates} is an equilibrium state, but it is actually a time-dependent, non-equilibrium state. In order to 
see this, we
need to consider correlators which include the operator $H$. Its special significance is that $\beta (H-E_0)$ acts as the modular Hamiltonian for the small algebra ${\cal A}$, which can detect
excitations in the commutant ${\cal A}'$ because of \eqref{comK}. 

Applying the same argument as in \eqref{proofeq} for correlators of the form $A H$ is not as straightforward because the typical size of $H$ is $O(S)$, hence it may enhance the $1/S$ corrections and lead to an $O(1)$ result. Indeed, as we will see below this actually happens and moreover these $O(1)$ terms can be reliably computed, provided that $\rsz$ has small energy spread.

Define $\hat{H}= H-E_0$ where $E_0 = \lsz H \rsz$. Including the normalization factor \eqref{normcorr} the state is $\rs = c V \rsz$. Here we introduce again the label $t_0$ to indicate that the unitary is made out of operators localized around $t=t_0$. Then we consider a correlator of the form
\be
\label{detectexc}
\langle \Psi| A(t) \hat{H} |\Psi\rangle 
\ee
We show that it is possible to select $A$ so that this correlator has nontrivial time-derivative with respect to $t$. The operator $A$ can be selected in several ways. One way is to first define the operator
$$
X(t_0) \equiv [H,U({\cal O}(t_0))]U^\dagger
$$
Notice that $X^\dagger=X$. Then we select 
\be
\label{bla3}
A(t) = e^{-{\beta H\over 2}} X(t) e^{\beta H \over 2}
\ee
We have
\be
\label{bor1}
{d \over dt}\ls A(t) \hat{H} \rs = |c|^2 {d \over dt}\lsz V^\dagger A(t) \hat{H} V\rsz  
\ee
\be
=|c|^2  {d \over dt} \lsz V^\dagger A(t) [H,V]\rsz + |c|^2 {d \over dt}\lsz V^\dagger A(t) V \hat{H}\rsz
\ee
The last term is subleading in $1/S$ and can be dropped. Intuitively
this is as follows: to leading order in $S^{-1}$ we have $\hat{H} \rsz = 0$. However, the state $\rsz$
will have some spread in energy, so there will be corrections. On the other hand from \eqref{bla4}
we have ${d \over dt}\lsz V^\dagger A(t) V \rsz = O(1/S)$. Combining these two statements we
arrive at the result mentioned above, see appendix \ref{estimateapp} for more details. Hence we have
\be
\label{bor2}
{d \over dt}\ls A(t) \hat{H} \rs = {d \over dt} \lsz V^\dagger A(t) [H,V]\rsz + O(S^{-1})
\ee
Now we replace in equation \eqref{bor2} that $[H,V] = e^{-{\beta H \over 2}}[H,U] e^{{\beta H \over 2}} =  e^{-{\beta H \over 2}}X(t_0) U e^{{\beta H \over 2}}$ and the expression \eqref{bla3} for $A$. Finally we approximate the RHS of \eqref{bor2} by the thermal trace
$$
{d \over dt}\ls A(t) \hat{H} \rs =  {d \over dt} Z^{-1}\Tr[e^{-\beta H} e^{\beta H\over 2}U^\dagger e^{-{\beta H \over 2}} e^{-{\beta H\over 2}} X(t) e^{\beta H \over 2} e^{-{\beta H \over 2}}X(t_0) U e^{{\beta H \over 2}}]+ O(S^{-1})
$$
Using the cyclicity of the trace
\be
\label{noneqfi}
{d \over dt}\ls A(t) \hat{H} \rs =  {d \over dt}Z^{-1}\Tr[e^{-\beta H} X(t) X(t_0)] + O(S^{-1})
\ee
where the leading term is generally nonzero and $O(S^0)$, showing that the state \eqref{newstates} has non-trivial time-dependence and is out of equilibrium.
\vskip10pt
\noindent {\bf Examples}
\vskip10pt
To be concrete, let us demonstrate the previous claims by considering the perturbation $U = e^{i \theta {\cal O}(t_0)}$ where ${\cal O}$ is a Hermitian operator. Ignoring the unimportant subleading normalization coefficient $c$,  the state is
\be
\label{statex}
\rs = e^{-{\beta H \over 2}} e^{i \theta {\cal O}(t_0)}e^{{\beta H \over 2}} \rsz
\ee
In this case we can simply select\footnote{The choice of $A(t)$ which can detect the excitation is not unique.} the operator $A$ to be $A(t)= {\cal O}(t)$ . First, to check \eqref{proofeq} we compute the expectation value of ${\cal O}(t)$ to linear order in $\theta$. We find
$$
\ls {\cal O}(t) \rs = \lsz e^{{\beta H \over 2}} e^{-i \theta {\cal O}(t_0)}e^{-{\beta H \over 2}}{\cal O}(t)  e^{-{\beta H \over 2}} e^{i \theta {\cal O}(t_0)}e^{{\beta H \over 2}}  \rsz
$$
$$
= \lsz {\cal O}(t) \rsz  + i \theta \left[\lsz {\cal O}(t) e^{-{\beta H \over 2}} {\cal O}(t_0) e^{{\beta H \over 2}}\rsz   - \lsz  e^{{\beta H \over 2}} {\cal O}(t_0) e^{-{\beta H \over 2}} {\cal O}(t)\rsz  \right]  + O (\theta^2)
$$
$$
= \lsz {\cal O}(t) \rsz + i \theta \left[\lsz {\cal O}(t)  {\cal O}(t_0 +i {\beta \over 2})\rsz   - \lsz   {\cal O}(t_0 -i {\beta \over 2})  {\cal O}(t)\rsz  \right]  + O (\theta^2)
$$
The first term is the equilibrium result, while the  linear term in $\theta$ in the brackets is zero by approximating these correlators by the thermal correlator and using the KMS condition for $\rsz$. Hence to this order the state \eqref{statex} seems to be in equilibrium. The same result holds if instead of ${\cal O}(t)$ we compute any other expectation value for an operator $A(t)$ from the algebra ${\cal A}$.

Second, to see that the state is out of equilibrium we  compute 
$$
\ls {\cal O}(t) \hat{H}\rs =
$$
$$
=\lsz {\cal O}(t) \hat{H}\rsz + i \theta \left[\lsz {\cal O}(t)  \hat{H} {\cal O}(t_0 +i {\beta \over 2})\rsz   - \lsz   {\cal O}(t_0 -i {\beta \over 2})  {\cal O}(t)  \hat{H}  \rsz  \right]  + O (\theta^2)
$$
$$
=\lsz {\cal O}(t) \hat{H}\rsz + i \theta \lsz {\cal O}(t)  [\hat{H}, {\cal O}](t_0 +i {\beta \over 2})\rsz
$$
$$
+i \theta  \lsz \left[ {\cal O}(t)   {\cal O}(t_0 +i {\beta \over 2})-  {\cal O}(t_0 -i {\beta \over 2})  {\cal O}(t)\right]  \hat{H}  \rsz    + O (\theta^2)
$$
The first term is the equilibrium result. The linear term in $\theta$ in the third line is very small by using the KMS condition and the fact that $\hat{H} \rsz \approx 0$, again the estimate for this type of correlator is in appendix \ref{estimateapp}. Hence the leading $O(1)$ term is
\be
\label{detectexb}
\ls {\cal O}(t) \hat{H}\rs =i \theta \lsz {\cal O}(t)  [H,{\cal O} ](t_0 +i {\beta \over 2})\rsz + O(S^{-1})
\ee
or
\be
\label{detectexb}
\ls {\cal O}(t) \hat{H}\rs = \theta Z^{-1} \Tr[{\cal O}(t)  {d{\cal O} \over dt}(t_0 +i {\beta \over 2})] + O(S^{-1})
\ee
This correlator decays exponentially as $|t-t_0|$ becomes large, but it is nonzero and $O(1)$ around the time $t=t_0$.

This confirms that the states \eqref{newstates} are out of equilibrium near $t=t_0$. They undergo a spontaneous fluctuation around that time and they settle back to equilibrium for earlier and later times. 

For example, consider a holographic 2d CFT on ${\mathbb S^1}\times {\rm time}$ in a typical pure state $\rsz$, dual to a BTZ black hole. We consider the Wightman 2-point function of a local scalar conformal primary operator of dimension $\Delta$, which in the large central charge limit is\footnote{According to \eqref{purevsmixedf}, we expect the large $c$ limit of the 2-point function on the typical pure state $\rsz$ to be close to the thermal correlator at the appropriate temperature, see also \cite{Fitzpatrick:2015zha} for a more direct demonstration of the same statement. The thermal real-time 2-point function on ${\mathbb S}^1 \times {\rm time}$ can be derived from that on ${\mathbb R}\times {\rm time}$ by using the method of images. For a general CFT the method of images cannot be applied since the operator ${\cal O}$ does not obey a Klein-Gordon equation on the boundary. However, this method can be applied in holographic CFTs in the high temperature phase, since the operator ${\cal O}$ is dual
to a bulk field $\phi$ which obeys the bulk Klein-Gordon equation and the spherical BTZ 2-point function can be derived from the planar BTZ 2-point function by the method of images. Finally in 2d the thermal 2-point function on ${\mathbb R}\times {\rm time}$ can be computed from that on the plane by the exponential map, see for example \cite{Papadodimas:2012aq}.}
\be
\label{btz2point}
\lsz {\cal O}(t,x) {\cal O}(t_0,x_0) \rsz = \left({2\pi \over \beta}\right)^{2\Delta}\sum_{m=-\infty}^{+\infty} {1 \over \left[2\cosh\left({2\pi (x-x_0 + 2\pi m) \over \beta} \right) 
- 2 \cosh\left({2\pi (t-t_0-i\epsilon ) \over \beta} \right) \right]^\Delta}
\ee
where the inverse temperature $\beta$ is determined by the energy of the state $\rsz$ using the equation of state of the CFT. The $i\epsilon$ fixes the phase when the operators are timelike. The sum over $m$ ensures the periodicity of the correlator along the spatial circle $x = x + 2\pi$.

\begin{figure}[!t]
\begin{center}
\includegraphics[width=.5\textwidth]{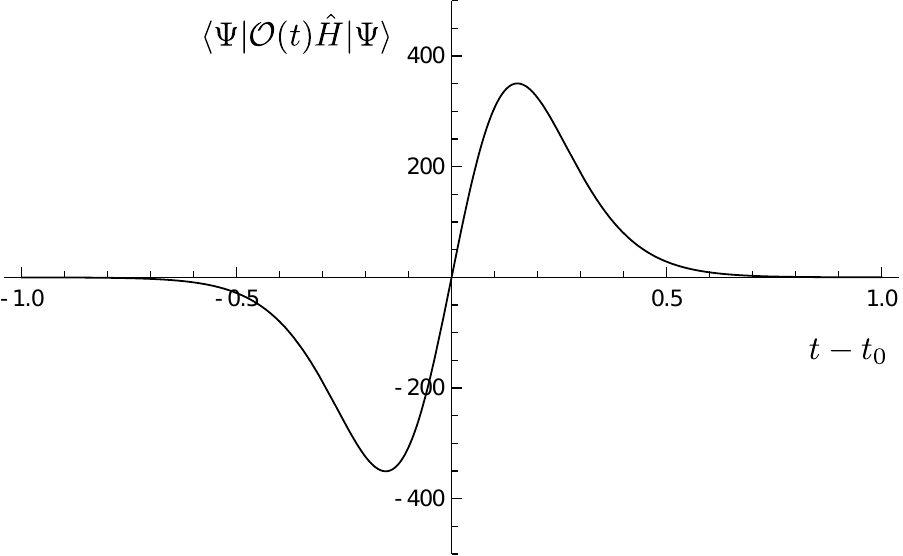}
\caption{The correlator \eqref{noneqcor} keeping only the linear term in $\theta$ and setting $\theta=1, \beta=1, \Delta=2$. The correlator is time dependent around $t=t_0$ and decays exponentially for large $|t-t_0|$.}
\label{fig3}
\end{center}
\end{figure}

Now, 
we place the two operators at different points in time, but the same location in space, and in what follows we suppress the $x$-variable. Analytically continuing $t_0\rightarrow t_0+i{\beta \over 2}$ we find
$$
\lsz {\cal O}(t) {\cal O}(t_0+i{\beta \over 2}) \rsz = \left({2\pi \over \beta}\right)^{2\Delta}\sum_{m=-\infty}^{+\infty} {1 \over \left[2\cosh\left({4\pi^2 m \over \beta} \right) 
+ 2 \cosh\left({2\pi (t-t_0) \over \beta} \right) \right]^\Delta}
$$
So if we consider the state \eqref{statex} and compute $\ls {\cal O}(t) \hat{H}\rs$ to linear order in $\theta$ using equation \eqref{detectexb} we find
\be
\label{noneqcor}
\ls {\cal O}(t) \hat{H}\rs =  \theta \, 2\Delta \left({2\pi \over \beta}\right)^{2\Delta+1} \sum_{m=-\infty}^{+\infty}{\sinh\left({2\pi (t-t_0) \over \beta}\right) \over \left[2\cosh\left({4\pi^2 m \over \beta} \right) 
+ 2 \cosh\left({2\pi (t-t_0) \over \beta} \right) \right]^{\Delta+1}}
\ee
In figure \ref{fig3} we plot the correlator as a function of $t-t_0$ for some choice of the values $\beta,\Delta$. 

While ${\cal O}(t)\hat{H}$ is not a Hermitian operator and thus not an observable, we could also consider the expectation value of the Hermitian operator $\{{\cal O}(t),\hat{H}\}$, which is 
twice the correlator \eqref{noneqcor}, given that the latter is real. Hence we find non-trivial time-dependence even in the expectation value of Hermitian physical observables.

\subsection{The bulk intepretation}

In AdS/CFT the algebra ${\cal A}$ is  generated by small products of single trace operators of low conformal dimension. In a theory with large $N$ and strong coupling these are dual to supergravity fields in the bulk. As mentioned in section \ref{sectherm} we do not include the CFT Hamiltonian $H$ in the algebra. Since we have
\be
H = \int_{S^{d-1}}d^{d-1}x\,\, T_{00}(t,x)
\ee
the zero mode of the stress tensor is not part of ${\cal A}$. This means that when we think of the stress tensor as a single trace operator dual to the perturbative graviton we need to subtract its zero mode. The fact that $H$ plays a special role and has to be treated differently is related to the fact that operators behind the horizon, if defined relationally with respect to the boundary, do not commute with the boundary Hamiltonian due to their gravitational dressing. This also comes out naturally from the 
algebraic point of view \eqref{comK}. Because of these non-vanishing commutators CFT correlators involving the Hamiltonian can detect the excitations by detecting the Gauss law tails. The values of these correlators is the same as what would be computed in effective field theory taking into account the gravitational dressing.

The states $e^{-{\beta H \over 2}} U({\cal O}(t_0))e^{{\beta H \over 2}}\rsz$ represent configurations which are undergoing a spontaneous fluctuation away from equilibrium around time $t=t_0$. This fluctuation cannot be detected by the usual HKLL fields in the exterior of the black hole. This suggests that these excitations live in a {\it causally disconnected region} in the bulk, which is naturally identified with 
part of the second asymptotic region. While these arguments do not constitute a proof that the black hole interior is smooth, they provide evidence in its favor. They show that the Hilbert space of the boundary CFT  contains states which can be naturally identified with excitations moving through the second asymptotic region that we would get by analytically continuing the solution past the horizons.

Let us also notice that if the state $U({\cal O})\rsz$ represents a wavepacket outside the horizon, then the conjugation by
the operator $e^{-\beta H/2}$ leads to the state $e^{-\beta H/2} U({\cal O}) e^{\beta H/2}\rsz$ which represents a similar (but conjugated) wavepacket, placed in the other asymptotic region behind the horizon. As discussed in section \ref{secrind} a similar observation holds for
Minkowksi space QFT. If the state $U({\cal O}_R)|0\rangle$ represents a wavepacket which is localized in the right Rindler wedge, then the state $e^{-\pi M} U({\cal O}_R) e^{\pi M}|0\rangle$, where now $M$ is the Lorentz boost generator, corresponds to a wavepacket localized in the left Rindler wedge.	Notice the structural similarity, since in the case of Minkowski space in the vacuum the Lorentz boost generator plays the role of the modular Hamiltonian, while for a typical equilibrium state the modular Hamiltonian is given by $\beta(H-E_0)$.

Finally, let us make a remark about the proposed interpretation of the states \eqref{newstates} as depicted in figure 1c. We see that the excitations of the states \eqref{newstates} move for a while in the left asymptotic region of the maximally extended Penrose diagram. On the other hand we do not expect a single CFT to be able to reconstruct the same spacetime as the one that would be assigned to the eternal black hole, which is holographically dual to two entangled CFTs. We expect that the reconstruction of the left region of the Penrose diagram for a 1-sided black hole in a typical state must be limited in some way, which we have not specified quantitatively. In the diagram this is indicated by the dotted lines, which mean that we cannot extend the spacetime all the way to the asymptotic region which would correspond to a second CFT. The fact that there is a limitation can be understood as follows: to approach the left asymptotic boundary we need to construct wavepackets which have very high energy. This means that the unitary $U({\cal O})$ must change the energy by a significant amount. In this limit the large $N$ expansion may become unreliable and some of the statements made above will have to be modified. Another way to see the limitation is that, since the factor $e^{-\beta H/2} U({\cal O}) e^{\beta H/2}\rsz$ lowers
the energy, the mass of the black hole provides a bound to how heavy the operator $U({\cal O})$ can possibly be. This indicates that there must be some effective cutoff in how much it makes sense to extend the diagram into the left asymptotic region. It would be very interesting to make this more precise.

\subsection{Some comments}
\label{seccomments}

\noindent a) The operator $e^{-{\beta H \over 2}} U({\cal O})e^{{\beta H \over 2}}$ in \eqref{newstates} is not unitary. However, for any given $|\Psi_0\rangle$, it is possible to rewrite the state $e^{-{\beta H \over 2}} U({\cal O})e^{{\beta H \over 2}} \rsz$ in the form $|\Psi\rangle = U(\widetilde{\cal O}) |\Psi_0\rangle$, 
where now $U(\widetilde{\cal O})$ is a unitary operator and the operators $\widetilde{\cal O}$ were defined in \eqref{defto}. This is analogous to \eqref{rindul} in the Rindler space discussion. However, in the case of the black hole the unitary $U(\widetilde{{\cal O}})$ is a state-dependent operator, i.e. it depends on the choice of the state $|\Psi_0\rangle$. Relatedly, $e^{-{\beta H \over 2}} U({\cal O})	 e^{{\beta H \over 2}} $ preserves the norm of the state only if it acts on equilibrium states, since in the derivation \eqref{proofnorm} we used the KMS condition for the state $\rsz$.

\vskip10pt

\noindent b) Unlike the states \eqref{standeq} discussed in the previous subsection which could be thought of as being prepared by a quench of the system, the states \eqref{newstates} are not so easy to prepare ''in practice``. This is 
related to the fact that the operator $e^{-{\beta H \over 2}} U({\cal O})e^{{\beta H \over 2}}$ is not a unitary. In principle we can produce them by a quench, where we perturb the Hamiltonian by the state-dependent operator $\widetilde{\cal O}$. Notice that in this context the use of state-dependent operators is conventional: unlike the infalling observer who has finite lifetime and resources, the boundary observer has unlimited resources. We could imagine a large number of systems prepared in an identical microstate. We use these copies to perform several measurements until the exact microstate is fully determined. Finally the specific operators $\tO$ relevant for this microstate can be applied to one of the (previously un-measured) copies of the system\footnote{We thank J. Maldacena for comments.}.

\vskip10pt

\noindent c) We emphasize once more that the states \eqref{newstates} definitely exist in the theory, if seen as ``autonomous states''. In other words, they are atypical states which undergo
a spontaneous deviation from equilibrium around the time $t=t_0$. 

\vskip10pt

\noindent d) We can also think of the states $e^{-{\beta H \over 2}} U({\cal O})e^{{\beta H \over 2}}\rsz$ as being prepared by starting with a Euclidean path integral to define the state. For example using the state-operator map in a CFT we start with the 
operator dual to state $\rsz$ at the center of ${\mathbb R}^d$, and then insert the unitary $U({\cal O})$ on the Euclidean plane, as determined by the Euclidean displacement operators $e^{\pm {\beta H \over 2}}$. In this was we get the desired state on the unit circle.

\begin{figure}[!t]
\begin{center}
\includegraphics[width=.6\textwidth]{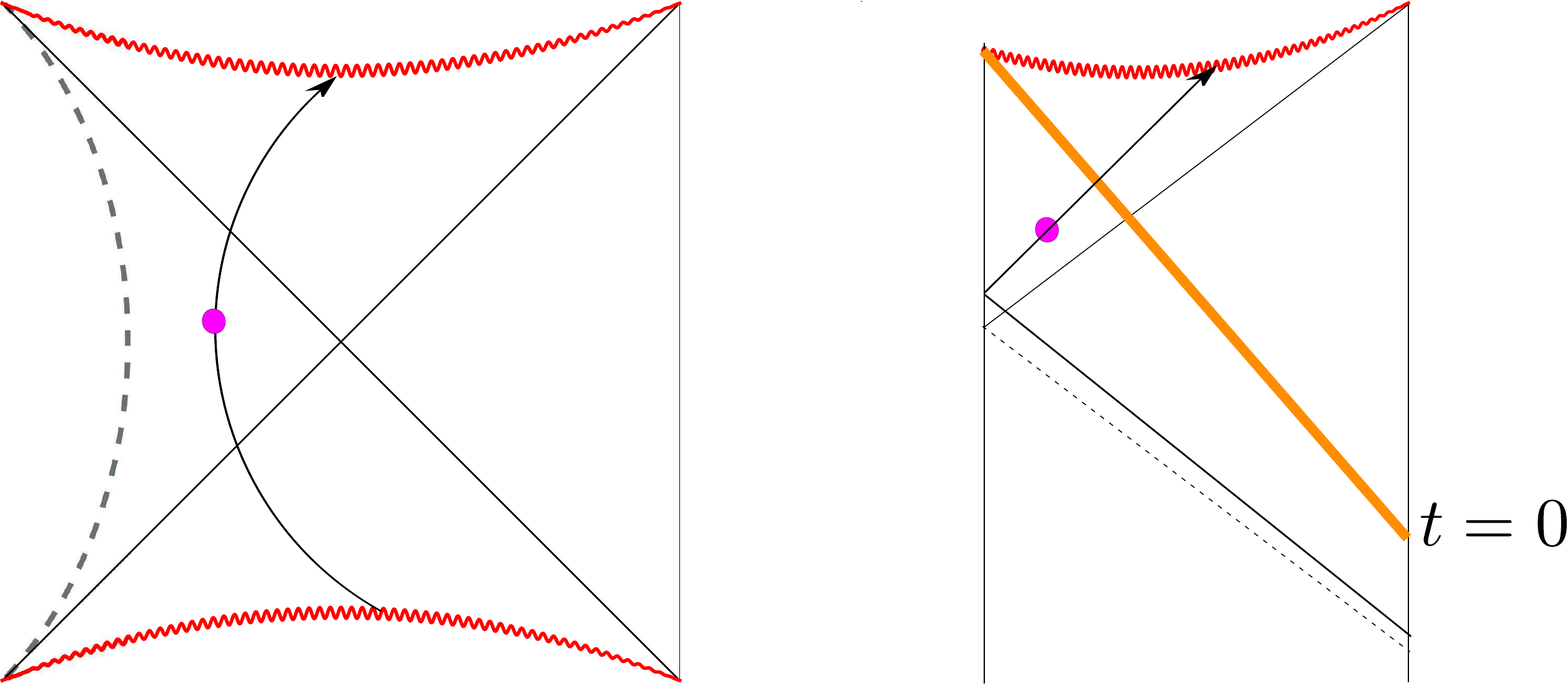}
\caption{Left: a non-equilibrium perturbation $e^{-{\beta H \over 2}}U({\cal O}) e^{{\beta H \over 2}}\rsz$ of a typical state $\rsz$. Right: a somewhat similar perturbation (purple particle) for a young black hole formed by the collapse of a shockwave emitted at $t=0$. The perturbation is created by dropping the particle before the shockwave. }
\label{figcoll}
\end{center}
\end{figure}

\vskip10pt

\noindent e)  We have discussed the non-equilibrium states of the form \eqref{newstates}, as small perturbations of typical states. Typical states are  different from black holes formed by gravitational collapse. The latter
correspond to special, atypical states. In a collapsing black hole background, excitations behind the horizon with some similarities to \eqref{newstates} can be created by sending particles from the exterior as indicated in figure \ref{figcoll}. These particles have to be dropped a little after the time where the extrapolation of the horizon to the past intersects the boundary as indicated in the figure. These infalling particles reach $r=0$ and start moving outwards. They go through the collapsing shell, and continue to move behind the horizon for a while until they fall into the singularity.  However, if we want to consider such right-moving particles behind the horizon of  the collapsing black hole at late times, we need to drop the particles with initially  transplanckian energy and thus effective field theory becomes unreliable. It would be interesting to understand better how to interpolate between the two regimes of typical vs collapsing black holes.

\subsection{On the energy of the non-equilibrium states}

It is interesting to compare the energy of the state 
\be
\label{newstateenergy}
|\Psi\rangle =c\, e^{-{\beta H\over 2}} U({\cal O}) e^{{\beta H \over 2}} |\Psi_0\rangle
\ee
to that of $\rsz$. As we will see the operator $ e^{-{\beta H\over 2}} U({\cal O}) e^{{\beta H \over 2}}$ decreases the expectation value of the energy, when acting on a typical state.

Before doing that, let us remind the reader that if we have a fixed unitary operator $U$ generated by the small algebra, then this operator will generally increase the energy of a typical state. 
To see that we want to compute the energy of the state
$$
U({\cal O}) \rsz
$$
compared to that of the typical state $\rsz$. We have
$$
\delta E = \lsz U^\dagger H U\rsz  - \lsz H \rsz
$$
$$
\delta E = \lsz U^\dagger [H,U] \rsz 
$$
Now since everything inside the correlator is an element of the small algebra, we can write
$$
\delta E = Z^{-1} \Tr[e^{-\beta H} U^\dagger [H,U]] + O(1/S) = Z^{-1} \Tr[e^{-\beta H} U^\dagger HU] - Z^{-1} \Tr[e^{-\beta H} H] + O(1/S)
$$
or
\be
\label{posen}
\delta E = \Tr[U \rho_\beta U^\dagger H] - \Tr[\rho_\beta H]  + O(1/S)
\ee
Ignoring the $1/S$ corrections, we can see that this expression is non-negative. The combination $\Tr[U \rho_\beta U^\dagger H] - \Tr[\rho_\beta H]$ is the change of the energy of a system in a thermal density matrix $\rho_\beta$. In that case it is impossible to lower the energy by acting on it with a unitary. One way to see it is by using the positivity of the relative entropy. The state we get after acting with the unitary
is $\rho' = U \rho_{\beta} U^\dagger$. By considering the positivity of the relative entropy $S(\rho'|\rho_\beta)$ and the fact that $\log(\rho_\beta) = - \beta H$, we find the inequality $\delta E \geq {\delta S \over \beta}$. However the two density matrices
$\rho',\rho_\beta$ have the same von Neumann entropy, hence $\delta S = 0$, from which we find that for states $U({\cal O})\rsz$ we have
\be
\delta E\geq 0
\ee 

Now let us repeat a similar argument for the state \eqref{newstateenergy}. We consider the difference of the expectation value of the Hamiltonian on the two states
$$
\delta E = \ls H \rs - \lsz H \rsz
$$
$$
\delta E = |c|^2 \lsz V^\dagger H V \rsz - \lsz H \rsz
$$
or
\be
\label{bla55}
\delta E = |c|^2 \lsz V^\dagger [H,V]\rsz + \left[|c|^2\lsz V^\dagger  V H\rsz - \lsz H \rsz \right]  
\ee
The second term is of order $1/S$ and we will concentrate on the first term. To see intuitively why the bracketed term  is subleading, consider the case where $\rsz$ has extremely highly peaked energy, for instance an exact energy eigenstate. Then we have that $H\rsz = E_0 \rsz$ and the term in the brackets is exactly zero, given how $c$ was defined \eqref{normconbb}. In the more general case where $\rsz$ has some spread in energy, provided that this spread is $O(T S^0)$ we can still show that the term in the bracket is $O(1/S)$, the  details can be found in appendix \ref{estimateapp}. We thus continue keeping only the first term in \eqref{bla55} and approximate it by a thermal correlator
\be
\label{lowen}
\delta E = Z^{-1} \Tr[e^{-\beta H} V^\dagger H V] - Z^{-1} \Tr[e^{-\beta H} V^\dagger V H] +O(S^{-1})
\ee
or
$$
\delta E = Z^{-1} \Tr[e^{-\beta H} e^{{\beta H \over 2}} U^\dagger e^{-{\beta H \over 2}}  H e^{-{\beta H \over 2}} U e^{{\beta H \over 2}}] - Z^{-1} \Tr[e^{-\beta H} e^{{\beta H \over 2}} U^\dagger e^{-{\beta H \over 2}} e^{-{\beta H \over 2}} U e^{{\beta H \over 2}} H]+O(S^{-1})
$$
or
$$
\delta E = Z^{-1} \Tr[e^{-\beta H} H] - Z^{-1} \Tr[U^\dagger e^{-\beta H}U H]+O(S^{-1})
$$
this has the opposite form of  \eqref{posen}, so for the states \eqref{newstateenergy} we find
\be
\delta E \leq 0
\ee
which is what we wanted to demonstrate.
\vskip30pt
\noindent {\bf Example}
\vskip10pt
\noindent Consider $U = e^{i \theta ({\cal O}_\omega + {\cal O}_\omega^\dagger)}$, where ${\cal O}_\omega = \int dt e^{i \omega t} {\cal O}(t)$. The first order term in $\theta$ is
\be
\delta E = i \theta \lsz \Big[ \hat{H} (e^{{\beta \omega \over 2}} {\cal O}_{\omega} + e^{-{\beta \omega \over 2}}
{\cal O}_\omega^\dagger)\Big]\rsz
  - i \theta \lsz \Big[  (e^{{\beta \omega \over 2}} {\cal O}_{\omega}^\dagger + e^{-{\beta \omega \over 2}}
{\cal O}_\omega)\hat{H}\Big]\rsz +O(\theta^2)
 \ee
 where $\hat{H} = H-E_0$. These terms are suppressed at large $S$ since the state $\rsz$ is an equilibrium state, see for example \eqref{eqcond},\eqref{eqcondb}. Moving on to quadratic order in $\theta$ and dropping some subleading terms we find
\begin{align}
\delta E =&\theta^2 \lsz \Big[e^{\beta \omega} {\cal O}_\omega^\dagger \hat{H} {\cal O}_\omega + e^{-\beta \omega} 
{\cal O}_\omega \hat{H} {\cal O}_\omega^\dagger -{1\over 2} \hat{H} ({\cal O}_\omega {\cal O}_\omega^\dagger + {\cal O}_\omega^\dagger {\cal O}_\omega)   -{1\over 2}  ({\cal O}_\omega {\cal O}_\omega^\dagger + {\cal O}_\omega^\dagger {\cal O}_\omega) \hat{H}\Big]\rsz
\end{align}
We can also write this as
\begin{align}
\label{quadensh}
& \delta E=\theta^2 \lsz \Big[e^{\beta \omega} {\cal O}_\omega^\dagger {\cal O}_\omega + e^{-\beta \omega} {\cal O}_\omega
{\cal O}_\omega^\dagger - {\cal O}_\omega {\cal O}_\omega^\dagger -{\cal O}_\omega^\dagger {\cal O}_\omega  \Big] \hat{H}
\rsz\cr
&- \theta^2  \omega \left[ e^{\beta \omega} \lsz {\cal O}_\omega^\dagger {\cal O}_\omega \rsz 
-e^{-\beta \omega} \lsz {\cal O}_\omega {\cal O}_\omega^\dagger \rsz \right]
\end{align}
After using the KMS condition we find up to quadratic order in $\theta$ and to leading order in $1/S$
\be
\label{lowere}
\delta E = -  \theta^2 \omega (e^{\beta \omega} -1) \lsz {\cal O}_\omega^\dagger {\cal O}_{\omega}\rsz  <0
\ee
Hence acting with the operator $e^{-{\beta H \over 2}} U({\cal O})e^{{\beta H \over 2}}$ lowers the expectation value of the energy of the state.

Notice that, had we done the same computation for the state $\rs = U({\cal O})\rsz$, we would have found an equation like \eqref{quadensh}, but without the factors $e^{\pm \beta \omega}$, which would imply 
$
\delta E =   \theta^2 \omega  (e^{\beta \omega} -1) \lsz {\cal O}_\omega^\dagger {\cal O}_{\omega}\rsz >0
$
\vskip10pt
\noindent {\bf Comments}

While the operator $e^{-{\beta H \over 2}} U({\cal O}) e^{{\beta H \over 2}}$ can decrease the energy of a {\it typical} state from the microcanonical ensemble, it is not unitary so we can not directly physically 
``act with it'' on a system, for example by adding a coupling to the Hamiltonian which is the same for all typical states. 
On the other hand, we can produce the state $e^{-{\beta H \over 2}} U({\cal O}) e^{{\beta H \over 2}}\rsz$  by the action of a  unitary on the equilibrium state $|\Psi_0\rangle$ which, as we explained, has to be a state-dependent operator $U(\widetilde{\cal O})$. This is a fine-tuned operator,  selected to lower the energy of {\it specifically} the state $|\Psi_0\rangle$. This situation, is perhaps somewhat analogous to the discussion about Maxwell's demon.

We argued that for typical states of given energy, the operator $e^{-{\beta H \over 2}} U({\cal O}) e^{{\beta H \over 2}}\rsz$ maps them to atypical states with lower expectation value of the energy. At first this might seem to be in contradiction with the fact that in general there are fewer states at lower energies than at higher energies\footnote{We would like to thank D. Jafferis, S. Minwalla and E. Rabinovici for discussions about this point.}. The reason that there is no contradiction is that while the states  $e^{-{\beta H \over 2}} U({\cal O}) e^{{\beta H \over 2}}\rsz$ have lower {\it expectation value} of the energy than the state $\rsz$, if we look at their expansion in energy eigenstates we notice that they also have contributions from eigenstates with energy larger than that of $\rsz$. In that sense they are using the larger Hilbert space at higher energies. The prefactor $e^{-{\beta H \over 2}}$ suppresses the amplitude of these higher energy contributions and enhances the low energy contributions, resulting in an overall lowering of the expectation value of the energy.

Relatedly, it is useful to consider the case where we start with the microcanonical density matrix $\rho_m = {{\mathbb P}_{E_0} \over {\cal N}}$ and we consider the transformation $V \rho_m V^\dagger$.\footnote{Notice that the canonical density matrix would formally be invariant, given that $(e^{-{\beta H \over 2}}U e^{\beta H \over 2})e^{-\beta H} (e^{{\beta H \over 2}} U^\dagger e^{-{\beta H \over 2}}) = e^{-\beta H}$, but we would have to be careful about the convergence of the operator $e^{\beta H \over 2}$ when it is acting on states with non-compact support in energy, as those relevant for the canonical ensemble. We thank J. Maldacena for related discussions.}  Given that $V$ is not unitary it is not obvious that this is an admissible
density matrix. However, let us define $\rho' = a V \rho_m V^\dagger$ where $a$ is a positive real number selected so that ${\rm Tr}[\rho']=1$. It has the form $a = 1 + O(1/S)$, which follows from $\Tr[V \rho_m V^\dagger] = Z^{-1}\Tr[\rho_\beta V^\dagger V] + O(1/S) = 1 +O(1/S)$. So the matrix $\rho'$ has unit trace, it is also Hermitian and positive, hence it is an admissible density matrix. Then we compute
$$
\delta E \equiv \Tr[\rho' H] -\Tr[\rho_m H]
$$
$$
= a \Tr[\rho_m V^\dagger [H,V]]  + \left[a \Tr[\rho_m V^\dagger V H] - \Tr[\rho_m H] \right]
$$
Assuming that the spread of the microcanonical is very small, the term in the brackets is approximately zero. Hence we find
$$
\delta E = a \Tr[\rho_m V^\dagger [H,V]]  
$$
Using that $a = 1 + O(1/S)$ and approximating this expression with the canonical ensemble we have
$$
\delta E =  \Tr[\rho_\beta V^\dagger [H,V]]  + O (1/S)
$$
Repeating the steps as in \eqref{lowen} we find
$$
\delta E = Z^{-1} \Tr[e^{-\beta H} H] - Z^{-1} \Tr[U^\dagger e^{-\beta H}U H]+O(S^{-1})
$$
which, as we explained above,  generally obeys
$$
\delta E \leq 0
$$
\section{Statistical aspects of  non-equilibrium states}

Here we discuss some statistical aspects of the non-equilibrium states \eqref{introstatesa}. The main points	 are the following. 
In equilibrium states correlators of observables are time-independent because of cancellations between random phases \eqref{ethexpf}. In standard non-equilibrium states of the form $U({\cal O})\rsz$ the phases 
are atypical and correlated, which leads to time-dependent signals for simple operators. In the non-equilibrium states of the form $e^{-{\beta H \over 2}} U({\cal O}) e^{\beta H \over 2} \rsz$ the phases are as correlated as in the states $U({\cal O})\rsz$, however
the insertion of the factors $e^{\pm {\beta H \over 2}}$ modulates the amplitude from  each energy bin, enhancing lower energies and suppressing higher energies. As a result there are cancellations between different energy bins, which are guaranteed by the KMS condition of the corresponding equilibrium state $\rsz$,  leading to time-independent correlators for the small algebra ${\cal A}$. Inserting factors of $H$ in the 
correlator spoils these cancellations between different energy bins
and reveals that the state is out of equilibrium.

This characterizes the qualitative property of the states \eqref{introstatesa}: they are atypical states where different energy bins are out of equilibrium, but when added together these bins conspire to disguise the state as an equilibrium state.
\subsection{Equilibrium states}

We consider a typical pure state $\rsz$ from the microcanonical ensemble, taking the energy spread to be highly peaked. 
In what follows we will assume that $\omega\neq 0$ and $\omega$ scales with entropy like $S^0$.
We expand the state as
$$
\rsz = \sum_i c_i |E_i\rangle
$$
where the coefficients $c_i$ are random and of typical size $e^{-S/2}$. We consider some observable $A$ from the small algebra ${\cal A}$ and define its Fourier transform $A_\omega = \int dt e^{i \omega t} A(t)$. From \eqref{ethdeff} we have
$$
(A_\omega)_{ij} = e^{-S/2} g(E,\omega) R_{ij}
$$
where $g(E, \omega)$ is a smooth real positive function of size $O(1)$ and $R_{ij}$ erratic phases. Here we use the notation $E=E_i$ and $\omega=E_j-E_i$. Hence\footnote{\label{smearomega}Since in general we assume that the energy gaps are non-degenerate \eqref{noncom}, one should smear the observables a bit in frequency space $A_\omega \rightarrow \int_{\omega-\delta\omega}^{\omega+\delta\omega} d\omega A_\omega$, so that a large
 number of pairs of energies $E_i,E_j$ will click. See \cite{Papadodimas:2013} for more details.} 
\begin{equation}
\label{eqcond}
\lsz A_\omega \rsz = \sum_{ij} c_i^* c_j R_{ij} e^{-S/2} g(E,\omega) = O(e^{-S/2})
\end{equation}

The smallness of this result is guaranteed by the random phases, and because of that it is quite robust. For example, it continues to be true when we include factors of $H$ in the correlator
\begin{equation}
\label{eqcondb}
\lsz A_\omega H \rsz = O(e^{-S/2})
\end{equation}
The phases of the state $H\rsz$ are the same as in $\rsz$, hence we have the same $e^{-S/2}$ suppression due to the erratic phases 
despite the fact that the operator $H$ scales like $S$. Notice that a naive approximation of this correlator via the canonical density matrix would give a less satisfying result
\be
\label{lessg}
\lsz A_\omega H \rsz = 0+ O(S^{-1})
\ee
where we used $ Z^{-1} {\rm Tr}[e^{-\beta H} A_\omega H] =0$ for $\omega \neq 0$ and the fact that generally correlators on $\rho_\beta$ and on typical states differ by $1/S$, \eqref{microvscanf} and \eqref{purevsmixedf} . The estimate \eqref{eqcondb} shows that the correlator has to be exponentially small, rather than just power-law suppressed as in \eqref{lessg}.

We remind the reader why not every expectation value on a typical state is exponentially small due to the random phases. For example we consider
\begin{align}
& \lsz A_\omega^\dagger A_\omega \rsz = \sum_{ijk} c_i^* c_k g^2(E,\omega) R_{ij}^\dagger R_{jk} e^{-S}\cr
& =\sum_{ij,k=i} |c_i|^2  g^2(E,\omega) |R_{ij}|^2 e^{-S} + O(e^{-S/2}) = O(1)
\end{align}
where in the first ``diagonal term'' we used that $|R_{ij}|^2=1$ and we are summing over $e^{2S}$ terms, each of which has typical size $e^{-2S}$ but now there are no erratic phases. Of course this is not
in contradiction with our criterion for equilibrium, as the operator $A_\omega^\dagger A_\omega$ has effectively zero frequency. It simply implies that in the correlator $\lsz A(t+t_1) A(t)\rsz$ there may be significant $t_1$ dependence, even if the
dependence on $t$ is exponentially small.

\subsection{Standard non-equilibrium states}

We now consider an ordinary non-equilibrium state of the form $\rs = U({\cal O}) \rsz$. In this and the next subsection we assume that the energy spread $\delta E$ of the underlying equilibrium  state $\rsz$ is very small. We expand it as
\begin{equation}
\label{coefnoneq}
\rs = \sum_i d_i |E_i\rangle 
\end{equation}
In this case we expect that there are observables for which $\ls A(t) \rs$ is time dependent, or equivalently that
$ \ls A_\omega \rs \neq 0$ even if $\omega\neq 0$. Using the matrix elements \eqref{ethdeff} we have
\be
\label{noneqsec5}
\ls A_\omega \rs = \sum_{ij} d_i^* d_j R_{ij} g(E,\omega) e^{-S/2}
\ee
If the coefficients $d_i$ were uncorrelated to the phases $R_{ij}$ this expectation value  would be $O(e^{-S/2})$. The fact that it is $O(1)$ means
that the $d_i$'s are correlated with $R_{ij}$. This is not surprising, since the unitary $U({\cal O})$ used to excite the state may be made out of the same operator ${\cal O}$ as the one used to detect the excitation. For example we can see this
in equation \eqref{noneqch}.
Moreover let us notice that under time evolution the coefficients evolve as $d_i \rightarrow e^{-i E_i t}d_i $ hence the phases decohere and the state equilibrates again.

We now consider the spectrum of the theory and we divide it into small energy bins of width $\delta E_{\rm bin}$. Here we take $\delta E_{\rm bin} = p S^0 T$, where $p\ll 1$. Hence $\delta E_{\rm bin}$ is much larger
than the typical energy gap $e^{-S} T$ and the energy bin $(E_a,E_a+\delta E_{\rm bin})$ contains an exponentially large number of microstates. At the same time the number of bins (the possible values of the index $a$) required to represent the states $\rs, \rsz$ scales like $S^0$, since we
assumed that the spread $|\Psi_0\rangle$ is very small and that the unitary $U$ changes the energy by an amount which scales like $S^0$.

Now, we define projection operators $\pP_{E_a}$ for each of these energy bins and we expand the non-equilibrium state $\rs = U({\cal O})\rsz$ as 
$$
\rs = \sum_a \pP_{E_a} \rs = \sum_a \rs_{a}
$$
where $\rs_{a}$ is the component of the state $\rs$ lying within the energy bin $(E_a,E_a+\delta E_{\rm bin})$.

We consider the expectation value of $A_\omega$ on the non-equilibrium state $\rs$ in the form
 \be
 \ls A_\omega \rs =  \sum_a   q_{\omega}(E_a)
\ee
where we defined
\be
\label{defineq}
q_{\omega}(E_a) \equiv \ls A_{\omega} \pP_{E_a} \rs = \lsz  U^\dagger A_{\omega} \pP_{E_a} U \rsz
\ee
as the contribution from each energy bin to the correlator. Expanding in energy eigenstates we have
\be
\label{bincont}
q_{\omega}(E_a) = \sum'_{i,j} d_i^* d_j g(E,\omega) R_{ij} e^{-S/2} 
\ee
where the prime in the sum means that we only sum over energy eigenstates which survive the projection to the bin imposed by ${\mathbb P}_{E_a}$. The sum over $j$ is directly constrained by the projector in \eqref{defineq}, while the sum over $i$ is constrained by the projector and by the shift
of the energy induced by $A_{\omega}$.

According to the ETH we would be tempted to say $q_{\omega}(E_a) \sim O (e^{-S/2})$, however for a non-equilibrium state the phases $d_i$ are correlated with $R_{ij}$. 
In fact, as we discussed above, the RHS of \eqref{noneqsec5} is nonzero and $O(S^0)$ in a non-equilibrium state. 
Given that we assumed that the number of bins does not scale like $S$, it is reasonable to assume that the contribution of each bin scales like  $q_{\omega}(E) = O(S^0)$ and is generally non-zero.

Since we get a signal of $O(S^0)$ from each bin, in some sense each of the energy bins is "out of equilibrium" and when we consider the entire state $\rs$ they all contribute to the correlator and give a nonzero result.

\subsection{The ``interior'' non-equilibrium states}
   We denote the equilibrium state as $\rsz$, the standard non-equilibrium states \eqref{standeq} as $\rs= U({\cal O})) \rsz$, and the state \eqref{newstates} as $\rsp =  e^{-{\beta H \over 2}} U({\cal O}) e^{{\beta H\over 2}}\rsz$. If the state $\rsz$ has highly peaked energy around $E_0$ (which we assume), then to leading order we can also write
$ \rsp=e^{-{\beta (H-E_0) \over 2}} U({\cal O}) \rsz$, which implies that $\rsp = e^{-{\beta (H-E_0) \over 2}} \rs$. If we expand this state as
$$
\rsp = \sum_i \td_i |E_i\rangle
$$
and we compare to \eqref{coefnoneq}, we notice that $\td_i = e^{-{\beta (E_i-E_0)\over 2}}d_i$. This means that the phases of $\td_i$ are the same as $d_i$ and only the magnitudes are modified by the energy dependent factor $e^{-{\beta (E-E_0) \over 2}}$. The fact that both $d_i$'s and $\widetilde{d_i}$'s correspond to an almost unit-normalized state was discussed in \eqref{proofnorm}.  We found before that for a non-equilibrium state we have $\ls A_\omega \rs = O(1)$ because the phases of $d_i$ are correlated with the phases $R_{ij}$ in the matrix elements of $A_\omega$. On the other hand
we found in  \eqref{proofeq} that correlators  of the algebra ${\cal A}$  on the state $\rsp = e^{-{\beta H\over 2}} U({\cal O}) e^{{\beta H \over 2}}\rsz$ are almost the same as thermal correlators, which implies that for $\omega \neq 0$ we have $\lsp A_\omega \rsp = 0$. This happens even though the phases of $\td_i$ are as correlated to $R_{ij}$ as those of $d_i$.

To see how this is possible we consider the contribution to the correlator on the state $\rsp$ from different energy bins
and we find
\begin{align}
& \lsp A_\omega \rsp =  \sum_{a}  \ls e^{-{\beta (H-E_0) \over 2}} A_\omega\, \pP_{E_a}\,\,e^{-{\beta (H-E_0) \over 2}}  \rs\cr 
& = \sum_{a}  e^{-\beta (E_a-E_0-{\omega\over 2})} \ls A_\omega \pP_{E_a} \rs\end{align}
or all in all
\begin{equation}
\label{cancelbin}
\lsp A_\omega \rsp = e^{\beta \omega \over 2} \sum_a e^{-\beta (E_a-E_0)} q_{\omega}(E_a)=0
\end{equation}
for the same $q_{\omega}(E_a)$ as those defined in \eqref{bincont}. While the sum is equal to zero, as guaranteed by \eqref{proofeq}, each of the $q_\omega(E)$ is generally nonzero and of $O(S^0)$ as we saw in \eqref{bincont}.

So we conclude that in the states of the form \eqref{newstates}, each of the microscopic energy bins is "out of equilibrium" but when we consider the contributions from all bins in the state we get cancellations between them  which make the state look like an equilibrium state. Inserting a factor of $H$ (or another function of $\pP_E$) in the correlator spoils these cancellations and in this way we can see the time-dependence as in \eqref{noneqfi}.

To see these cancellations in an example, suppose we consider the operator $A_\omega={\cal O}_\omega$ and the state 
$$U = e^{-{\beta H \over 2}}e^{i \theta (\co_\omega + \co^\dagger_\omega)}e^{{\beta H \over 2}}\rsz$$ and expand to linear order in $\theta$. We find
$$
q_\omega(E) = \lsz \co_\omega {\mathbb P}_E \rsz + i\theta \left[\lsz \co_\omega {\mathbb P}_E (\co_\omega + \co^\dagger_\omega)\rsz-
\lsz (\co_\omega + \co^\dagger_\omega)\co_\omega {\mathbb P}_E \rsz\right]
$$
The first term is zero, so we find
$$
q_\omega(E) = i\theta \left[\lsz \co_\omega {\mathbb P}_E  \co^\dagger_\omega\rsz-\lsz  \co^\dagger_\omega\co_\omega {\mathbb P}_E \rsz\right] 
$$
Approximating the division over energy bins by a continuous distribution and using the KMS condition
$$
q_\omega(E) = i \theta \lsz {\cal O}_\omega^\dagger {\cal O}_\omega\rsz \left[e^{\beta \omega}\delta(E - E_0-\omega) - \delta(E-E_0) \right] 
$$
Computing the expectation value of $\co_\omega$ on the state $\rs=U\rsz$ we find
$$
\ls \co_\omega \rs = \int dE q(E) = i \theta \lsz{\cal O}_\omega^\dagger {\cal O}_\omega\rsz (e^{\beta \omega} - 1)
$$
On the other hand for the state $\rsp = e^{-{\beta H \over 2}} U e^{{\beta H \over 2}} \rsz$ we find
$$
\lsp \co_\omega \rsp = \int dE e^{-\beta (E-E_0)} q_\omega(E) = i \theta \lsz{\cal O}_\omega^\dagger {\cal O}_\omega\rsz (e^{-\beta \omega} \times e^{\beta \omega} - 1) = 0
$$
which is what we expected. We emphasize that this cancellation worked out because of the KMS condition of the underlying equilibrium state.

\section{Discussion}

We considered a class of non-equilibrium states, which are generally present in any quantum system. Independent of the discussion about the black hole interior, these states represent a mathematically canonical class of non-equilibrium states. It would be interesting to explore further what is their role in the process of thermalization of a system in a pure state.

In AdS/CFT, and if we accept that the black hole interior is smooth for typical states, these non-equilibrium states describe black holes with excitations behind the horizon. The existence
of these states in the CFT is an indication in favor of the smoothness of the interior. However, it is not a proof by itself, since in order to verify how the infalling observer experiences the states we need to construct operators behind the horizon.
The proposal of \cite{Papadodimas:2012aq,Papadodimas:2013b,Papadodimas:2013,Papadodimas:2013kwa,Papadodimas:2015xma,Papadodimas:2015jra}, when applied to these states, reproduces the spacetime diagram of figure \ref{fig1}.

In this paper we only discussed states where the  perturbation by the unitary $U$ changes the energy by order $O(N^0)$, which means that backreaction to the classical geometry is negligible. It would be interesting to understand how these states start to behave once they become heavy enough to backreact on the geometry.

 In this paper we considered excitations of pure states $\rsz$ corresponding to the microcanonical ensemble with small energy width. It would be interesting to investigate excitations of more general states. 
 \vskip20pt
 \begin{acknowledgments}
  I would like to thank C. Bachas, S. Banerjee, J. Barbon, R. van Breukelen, J.W. Bryan, I. Bena,  J. de Boer, M. Floratos, M. Guica, D. Harlow, D. Jafferis, E. Kiritsis,  J. Maldacena, D. Marolf, S. Minwalla, L. Motl, J. Penedones, J. Polchinski, E. Rabinovici, K. Skenderis, D. Turton,  E. Verlinde, G. Vos and S. Zhiboedov for discussions over the last few years.
  I especially thank S. Raju for discussions and collaboration on related topics, and J. Barbon and E. Rabinovici for comments on the draft. I am especially grateful to J. Maldacena for correspondence and comments on the draft. I would like to thank ENS, Paris and IHES  for hospitality during completion of this work and the Royal Netherlands Academy of Sciences (KNAW).
 \end{acknowledgments}
\vskip20pt
\appendix
\addtocontents{toc}{\protect\setcounter{tocdepth}{1}}
\addtocontents{lof}{\protect\setcounter{tocdepth}{1}}

\section{Time dependence, equilibrium states, spontaneous fluctuations}
\label{statmech}

This is a more expanded version of the discussion in section \ref{eqmaintext}. This appendix contains mainly well known results.

In order to study the time-dependence and equilibration of a quantum statistical system we will make some assumptions about the spectrum and the observables. Regarding the spectrum, we consider a closed, bounded quantum system with a discrete energy spectrum $E_i$. We will assume that the energy eigenvalues $E_i$ are non-degenerate, the energy gaps $E_i-E_j$ are non-degenerate and more generally that for any finite number $m$ of energy eigenvalues $E_i$, any linear combination with integer coefficients $n_i$ obeys
\be
\label{noncom}
\sum_{i=1}^m n_i E_i =0 \qquad \Leftrightarrow \qquad n_i=0
\ee
In practice, this does not need to be true for all states. For example spacetime and global symmetries imply that there must be degeneracies, or that the energy gaps may be degenerate. However these degeneracies relate a small number of states relative to the total number of states at a given energy range, and this does not affect the main conclusions significantly. While QFTs  have infinite-dimensional Hilbert spaces, for most questions we want to study, only a finite number of states play an important role.

Even though we consider systems with fundamentally discrete spectrum, in many cases the density of states is so high that we can approximate it by a continuous spectrum. For example, a large $N$ gauge theory defined on a compact spatial manifold has discrete spectrum. However, in the high-temperature phase and in the 't Hooft large $N$ limit we find that the density of states scales like $\beta e^{s(\beta) N^2}$ where $s(\beta)$ is some $N$-independent function of the temperature, or the energy gap between energy eigenstates is of the order $ \beta^{-1} e^{-s(\beta) N^2}$. Thus in the large $N$ limit the spectrum becomes almost continuous, which is important in order to reproduce expected properties of thermalization dual to black hole behavior in AdS/CFT \cite{Festuccia:2006sa}.

Moving to the observables, we consider a class which we think of as ``simple'' or ``coarse-grained observables''. They have the following properties: any Hermitian observable can be written
in a spectral decomposition
$$
A = \sum_i   a_i  {\mathbb P}_i
$$
where $a_i$ are the eigenvalues of $A$ and ${\mathbb P}_i$ is a projector on the eigenspace of eigenvalue $a_i$. We will consider observables where the number of different eigenvalues is parametrically smaller than the dimensionality of the Hilbert space ${\cal N}=e^{S}$. This means that each of
the projectors has dimensionality which is comparable to ${\cal N}$.

The second assumption about the observables is that the subspaces ${\mathbb P}_i$ are ''randomly-oriented`` with respect to the energy eigenstates. This is related to the Eigenstate Thermalization Hypothesis (ETH) \cite{peresETH, deutsch, srednicki1999approach}, which postulates that the matrix
elements of observables on energy eigenstates have the following structure
\be
\label{ethdef}
\langle E_i| A |E_j\rangle = f(E_i) \delta_{ij} + R_{ij} e^{-S/2} g(E_i,E_j)
\ee
where the functions $f,g$ are smooth functions of the energy and of magnitude $O(S^0)$, while $R_{ij}$ are complex coefficients with magnitude of $O(1)$ and erratic phase. We will assume that the observables we consider obey \eqref{ethdef}.

Finally, we want to remind how we compare different quantum states. For two unit-normalized pure states, we can consider  their inner product
\be
\label{diststates}
|\langle \Psi| \Psi'\rangle| =\sqrt{1-\epsilon^2} 
\ee
where $0\leq\epsilon\leq 1$. As $\epsilon\rightarrow 0$ the two states start to look the same. More specifically, for any bounded operator $A$ the difference of the expectation value of $A$ on these two states is bounded
by the operator norm of $A$ and the distance between the states characterized via \eqref{diststates}, by the equation
\be
\label{distobs}
| \ls A \rs - \lsp A \rsp|\leq\,||A|| \,\epsilon 
\ee
This also applies to projectors, which have norm $||{\mathbb P}||=1$, and in particular it implies that if two state-vectors are close enough in the Hilbert space norm, then the prediction of quantum probabilities on the two states given by the Born rule will also be close.

\subsection{Comments on time dependence in quantum mechanics}

The time evolution of a state in the Schroedinger picture is given by
\be
\label{qmev}
|\Psi(t) \rangle = \sum_i c_i e^{-i E_i t} |E_i\rangle
\ee
Assuming that the spectrum is chaotic, in the sense of  \eqref{noncom}, time evolution randomizes the phases but the distribution of the magnitudes $|c_i|$ remains constant throughout the history of the state. So if we start with a 
state with some unusual pattern in the distribution of magnitudes of $|c_i|$, this pattern will be preserved in time.

To partly characterize the distribution of magnitudes $|c_i|$  in the superposition describing a pure state we define 
$$
 E\equiv \ls H \rs
$$
and 
\be
\label{enspread}
(\Delta E)^2 \equiv  \ls H^2 \rs - \ls H \rs^2
\ee
Given the form of time evolution \eqref{qmev} these two quantities remain constant throughout the history of a quantum state.

We now want to study the time-evolution of observables
$$
\langle A(t)\rangle = \langle \Psi(t)| A |\Psi(t)\rangle
$$
Notice that $A$ may be a product of several operators, which can even be displaced in time. In that case $t$ represents the overall time-shift of the product of  operators making up $A$.

If we want the observables to have nontrivial time dependence on a state, then the energy spread \eqref{enspread} must be large enough. For example, in an exact energy eigenstate we have $\Delta E=0$ and the
expectation values of all observables are exactly time-independent. More generally, if we consider the variance of the observable  $\Delta A^2 \equiv \langle \Psi| A^2 |\Psi\rangle - \langle \Psi| A |\Psi\rangle^2$ then we have
\be
\label{dedt}
{1\over 2}\left|{d \langle A\rangle  \over dt} \right|\leq \Delta A\, \Delta E
\ee
This basic inequality shows that if we want to consider states with interesting time-dependence, then they must include a relatively broad range of energy eigenstates in the superposition defining the state $\rs = \sum_i c_i |E_i\rangle$. For instance, suppose we
have a big black hole in AdS and a particle falling towards the horizon. This state has the property that single-trace correlators have time-dependence over a time scale which is of the order
of the inverse temperature, which means that ${d \langle A\rangle  \over dt}$ scales like $N^0$. Also for such single trace operators (appropriately smeared to avoid UV divergences) we have $\Delta A = O(N^0)$. 
Hence such a state must have at least $\Delta E = O(T N^0)$.

Conversely, if we take a state where $\Delta E$ is parametrically smaller, for example if it scales like some positive power of $1/N$, then by \eqref{dedt} it is guaranteed that we will not get time-dependent signals in single trace operators\footnote{Here we refer to the expectation value of a given operator $A$, which itself may be a product of various local operators separated in time, for instance applying \eqref{dedt} to $A= e^{iHt_1} {\cal O}(0) e^{-i H t_1}{\cal O}(0)$ would only bound the dependence on $t$ and not on $t_1$.} with magnitude of $O(1)$,  even if we wait an exponentially long time.

The inequality \eqref{dedt} is also relevant for the following point: sometimes it is stated that in AdS/CFT we can produce a black hole in the microcanonical ensemble (with small energy spread) by collapsing a shell of matter with that energy. Then \eqref{dedt} demonstrates that this is impossible: in a collapsing shell geometry even the 1-point functions (for example of HKLL-like operators) have non-trivial time dependence over time scales of order $N^0$. These 1-point functions correspond to classical vevs\footnote{This is in conventions where the 1-point functions of single trace operators on a semiclassical bulk solution are of order $N^0$.} for which $\Delta A \sim O(1/N)$. Thus in order to have nontrivial time dependence we need $\Delta E \sim O(N)$ and it is impossible to form a black hole whose energy spread is small enough for the state to belong to the microcanonical ensemble.

\vskip10pt

\noindent {\bf Recurrences}

A pure state in a finite quantum system undergoes Poincare recurrences. This means that the system will look like the initial state $|\Psi\rangle$ an infinite number of times. We can study this  either in terms of correlators, or in terms of the proximity of the state $|\Psi(t)\rangle$ to $|\Psi(0)\rangle$. For example, consider the inner product
$$
h(t) \equiv \langle \Psi (0) | \Psi(t)\rangle = \sum_i |c_i|^2 e^{-i E_i t} 
$$
Here we can assume that the sum over $i$ is finite, over $e^{S}$ states\footnote{Even if the state $|\Psi(0)\rangle$ is a superposition of an infinite number of energy eigenstates, if the state is unit-normalized then we can truncate the sum to a finite superposition without affecting the magnitude of the inner product significantly.}. We have $h(0)=1$. As $t$ increases $h(t)$ decreases down to a size of order $e^{-S/2}$, around which it continuous to fluctuate. At time scales of order $e^{e^S}$ there will be moments when $h(t) \approx 1$. At those moments the states $|\Psi\rangle$ and $|\Psi(t)\rangle$ will look almost the same for most observables \eqref{distobs}. Hence the system comes back to itself.

For an energy eigenstate there is no recurrence, or equivalently the recurrence time is zero. For small energy spread $\Delta E$ we do not have interesting Poincare recurrences, since equation \eqref{dedt} implies that all reasonable observables look almost static ---
though it would be interesting to explore whether there are any interesting ''slow observables`` to consider. For  discussions about recurrences in the context of black holes in AdS/CFT see \cite{Maldacena:2001kr,Barbon:2003aq,Barbon:2014rma,Fitzpatrick:2016mjq}.

\subsection{Ensembles and typical pure states}
\label{mixedpure}

When studying an isolated system in a pure state, it is useful to define the notion of a {\it typical pure state}. This requires the definition of an ''ensemble`` of pure states, with some particular probability measure defined for them.

One ensemble is the microcanonical ensemble. We specify an energy $E_0$ and a spread of energy $\delta E$. We consider all energy eigenstates $|E_i\rangle$ contained in this energy window. We assume that the spectrum is discrete, so there is a finite number ${\cal N} = e^{S}$ of energy eigenstates in that window. We write down the most general superposition of those eigenstates
\be
\label{micropure}
|\Psi\rangle = \sum_i c_i |E_i\rangle
\ee
The coefficients must satisfy $\sum_i |c_i|^2 = 1$. So every pure state-vector is related to a point on the ${\mathbb S}^{2{\cal N}-1}$ sphere --- the fact that physical states are rays is not important here. The microcanonical measure $d\mu$ is defined by assigning a 
probability distribution on the sphere which is proportional to the volume element of the round sphere. It is equivalent to starting with one particular state on this space, and defining a measure by acting on it with all possible unitaries selected with the $U({\cal N})$ Haar measure.

Using this microcanonical measure we can define averages over all possible pure states of the form \eqref{micropure}. For example, we define
\be
\overline{\ls A \rs} \equiv \int d\mu \ls A \rs
\ee
Whenever we make a statement about a  typical state, we mean that the statement is true for the largest volume of pure states on this sphere.

There is also a corresponding mixed state 
\be
\label{mixedmicro}
\rho_m \equiv {{\mathbb P}_{E_0} \over {\cal N}}
\ee
where ${\mathbb P}_{E_0}$ is the projector on the Hilbert subspace spanned by eigenstates in the given window of energies.

Now, consider an observable $A$. It is useful to consider how much the expectation value of $A$ varies among the different pure states \eqref{micropure} and how much it differs from the corresponding expectation value on the mixed state \eqref{mixedmicro}. It is easy to show that
\be
\label{lloyda}
\overline{\ls A \rs} =\Tr[\rho_m A]
\ee
Moreover we can compute the variance over different pure states from the microcanonical ensemble. For simplicity we take $A$ to be Hermitian. We have
\be
\label{lloydb}
\overline{ \left(\ls A \rs - \Tr[\rho_m A] \right)^2} = {1\over e^S+1} \left(\Tr[\rho_m A {\mathbb P}_{E_0} A] - \Tr[\rho_m A]^2 \right)
\ee
which implies
\be
\label{lloydc}
\overline{ \left(\ls A \rs - \Tr[\rho_m A] \right)^2} \leq {1\over e^S+1} \left(\Tr[\rho_m A^2] - \Tr[\rho_m A]^2 \right)
\ee
This fundamental result has been discussed in various places \cite{lloyd,Balasubramanian:2007qv, Lashkari:2014pna, Harlow:2014yoa}. Notice that equation \eqref{lloydb} is an identity, which does not depend on the complexity of the observable $A$ or on the Hamiltonian of the system we are considering.

If we work with operators $A$ whose operator norm is bounded and of $O(S^0)$, then we notice that the RHS of \eqref{lloydb} is exponentially suppressed in the entropy. For instance, if we probe the state by using projectors which have $||{\mathbb P}||=1$, then the variance is always exponentially suppressed.  From \eqref{lloydb} it is clear that if we want to identify whether the system is in a  particular microstate, then we need to measure an observable which is fine-tuned relative to the state that we want to detect. For instance, we could use the projector $\rs\ls$ corresponding to the state in question. Conversely, reasonable coarse-grained observables cannot effectively distinguish among different pure states, or between pure states and the microcanonical mixed state \eqref{mixedmicro}. For such observables and for typical states of the form \eqref{micropure} we have
\be
\label{purevsmixed}
\ls A \rs = \Tr[\rho_m A] + O(e^{-S/2})
\ee

In the case of the microcanonical ensemble there is a clear connection between the maximally mixed density matrix $\rho_m$ and the probability measure $d\mu$ on the set of pure states that we defined above. One may ask whether there are other interesting probability measures for pure states. For example, we can ask if there is a measure that can be defined on pure states, with
the property that typical pure states with respect to that measure give correlators which are exponentially close to those in the mixed state $\rho_\beta \equiv {e^{-\beta H} \over Z}$. For the purpose of studying equilibration we want to find measures which are invariant under time evolution \eqref{qmev}. An extreme choice would be to consider an ensemble of pure states which are superpositions of all energy eigenstates, with fixed magnitudes of the coefficients $c_i$
to be $|c_i|^2 = {e^{-\beta E_i } \over Z}$ and allow the phases to vary, assigning uniform probability to each phase. It would however be more natural to allow the magnitudes of the coefficients to vary as well. We will not address this question further in this paper, but for some related discussions  see \cite{Papadodimas:2015jra}.

\subsection{Equilibrium states}
\label{eqreview}

We now want to define the notion of an equilibrium {\it pure} state. As we discussed in section \ref{eqmaintext} the two conditions we will require are: i) that correlators of observables on the state are time-independent and ii) that their values are close to the thermal correlators on $\rho_\beta$, or equivalently up to $1/S$ corrections, to microcanonical correlators on $\rho_m$.\footnote{The second condition is useful
in order to exclude certain states, for example consider the state $|\Psi\rangle = {1\over \sqrt{2}}(|E_1\rangle + |E_2\rangle)$, where $|E_{1,2}\rangle$ are two energy eigenstates with widely different energy. According to the ETH \eqref{ethdef} correlators on $\rs$ are approximately time-independent, but far from those of the canonical, or microcanonical ensembles.}

However, when we consider pure states then generally correlators of operators cannot be exactly time independent, unless the pure state is an energy eigenstate. Energy eigenstates are exactly static, and moreover the ETH \eqref{ethdef} implies that correlators on energy eigenstates give the same values (up to $1/S$ corrections) to thermal correlators\footnote{This statement is true for ''most`` energy eigenstates.}. Hence an energy eigenstate can be classified as a special case of an equilibrium state. 

For pure states with nonzero energy spread $\delta E$, the observables will have non-trivial time dependence. It is useful to consider the long-time average \cite{srednicki1999approach}, defined by
\begin{align}
\overline{A_t} \equiv \lim_{T\rightarrow \infty} {1\over T}\int_0^T dt \langle \Psi(t)|A|\Psi(t)\rangle  &= \lim_{T\rightarrow \infty} {1\over T}\int_0^T dt \sum_{ij} c_j^* c_i A_{ji} e^{-i(E_i-E_j)t}\cr&  = \sum_i |c_i|^2 A_{ii} = \Tr[\rho_\Psi A]
\end{align}
where we used the assumption \eqref{noncom} and we defined
\be
\label{longteq}
\rho_{\Psi} \equiv \sum_i |c_i|^2 |E_i\rangle \langle E_i|
\ee

The density matrix \eqref{longteq} characterizes the long-time equilibrium mixed state corresponding to the pure state $\rs$. We also consider the fluctuations of the expectation value of the observable around the long time average. We find
\begin{align}
\label{longtvar}
 \overline{(A_t- \overline{A_t})^2} = \lim_{T\rightarrow \infty}{1\over T}  \int_0^T dt (\langle A\rangle_t - \overline{A_t})^2  =
 \sum_{i\neq j} |c_i|^2 |c_j|^2 |\langle i| A |j \rangle|^2
\end{align}
If most of the coefficients $c_i$ within the energy band $\delta E$ are approximately equal then $|c_i|^2 \sim O(e^{-S})$. Also according to the ETH the matrix element $|\langle i |A|j \rangle|$ is of order $O(e^{-S/2})$ if $i\neq j$, and we are summing over $e^{2S}$ terms. Putting everything together the RHS of \eqref{longtvar} is of order $e^{-S}$. This shows that for the vast majority of time the pure state $|\Psi(t)\rangle$ looks like the long-time average equilibrium state \eqref{longteq}, which is indeed time-independent. Also, from the ETH, if we assume that the function $f$ in \eqref{ethdef} is slowly varying, then correlators on $\rho_{\Psi}$ are almost the same as thermal correlators. Hence for most time a pure state will look like an equilibrium state.

We can also see that pure states appear to be  time-independent for most of the time as follows. Using the expansion \eqref{ethdef}, for any state we have
\be
\label{ethexp}
\langle A \rangle = \sum_i |c_i|^2 f(E_i)  + \sum_{i\neq j} e^{-i (E_j-E_i)t} c_i^* c_j R_{ij}e^{-S/2} g(E_i,E_j) 
\ee
and for the time derivative
\be
\label{randomphase}
{d \langle A \rangle \over dt} = -i\sum_{ij} (E_j-E_i) e^{-i(E_j-E_i)t} c_i^* c_j R_{ij}e^{-S/2} g(E_i,E_j) 
\ee
For typical pure states, as defined in the previous subsection, the phases of the coefficients $c_i$ are randomly distributed, and in particular they are uncorrelated with the 
erratic phases $R_{ij}$ of the matrix element $A_{ij}$. Hence the RHS of this expression is of order $e^{-S/2}$, which shows that observables are almost time-independent on
typical states.

Above we argued that the variance of the expectation value of $A$ around the long-time equilibrium value $\overline{A_t}$ is exponentially small in the entropy. Of course at any given moment in time, we may have significant quantum/thermal variance regarding the outcomes of a measurement of the observable $A$. This quantum variance at time $t$ is controlled by the quantity $\langle \Psi(t)| A^2|\Psi(t) \rangle - \langle \Psi(t)| A |\Psi(t)\rangle^2$, which is different from the long-time variance of the expectation value \eqref{longtvar}. 

Relatedly, in the main text we discussed spontaneous fluctuations of pure states, due to ''resonances`` of the time-dependent off-diagonal phase factors in \eqref{ethexp}. We emphasize that these spontaneous fluctuations of pure states have to be distinguished from the fact that observables have nontrivial quantum/thermal variance. For example, in the thermal density matrix $\rho_{\beta} = Z^{-1} e^{-\beta H}$, observables
generally have a nonzero variance, which means that the outcomes of various measurements follow a probability distribution which allows certain unlikely outcomes to happen with nonzero, but small probability. The spontaneous fluctuations that we are considering are  not of this type. When a pure state undergoes a spontaneous fluctuation due to the resonance between the phase factors $e^{-i(E_j-E_i)t} c_i^* c_j $ and $R_{ij}$, what happens is that the {\it expectation values} of observables deviate from the equilibrium values, i.e. the probability distribution for various outcomes is modified relative to that in the thermal ensemble.

Finally let us remind the reader that one has to be careful when considering the set of all equilibrium states. The first point is that, as explained above, whether a state is equilibrium or not may depend on the time that we probe it. So we should be talking about the set of equilibrium states at a given time range. The second point is that the set of equilibrium states does not form a vector space. For example, according to our previous definitions, energy eigenstates are equilibrium states, but clearly by superimposing them we can get arbitrary non-equilibrium states. Conversely, by superimposing a large number of non-equilibrium states we can produce an equilibrium state. While the notion of equilibrium is not preserved under arbitrary linear combinations, it is generally preserved under the superposition of a small number of states. More details about this point can be found in \cite{Papadodimas:2015jra}.

\section{On the domain of complexified boosts}
\label{domainboost}

We consider the Lorentz boost generator $M$ on the $t$-$x^1$ plane, which also corresponds to time-translations in Rindler time. We want to understand that states of the form ${\cal A}_R |0\rangle$ are in the domain 
of the complexified boost $e^{-s M}$ for $0\leq s \leq \pi$. In general, states of the form ${\cal A}_L |0\rangle$ are {\it not} in the same domain, but rather in the domain of complexified boosts with $-\pi \leq s \leq 0$. Consider a local scalar operator ${\cal O}(x)$ and the smeared operator 
$$
{\cal O}(f) \equiv \int d^dx f(x) {\cal O}(x)
$$
where $f$ is smooth  and has support on the right wedge. We assume that we have specified a prescription of the smeared operator which leads to a normalizable Lorentzian state
$$
|\Psi\rangle = {\cal O}(f) |0\rangle
$$
This state is in the domain of the real Lorentz boost $e^{i \tau M}$ for all $\tau$. We want to complexify $\tau$. In general we have
$$
e^{i\tau M} |\Psi\rangle = \int d^dx f(x)   e^{i\tau M} {\cal O}(x) \vac = \int d^d x f(x) e^{i\tau M} e^{i P_\mu x^\mu} {\cal O}(0) e^{-i P_\mu x^\mu} \vac
$$
Now we complexify $\tau \rightarrow i s$ with $s\in {\mathbb R}$ and use the Poincare algebra to write
$$
e^{ -s M} e^{i P_\mu x^\mu} = e^{i \cos s P_\mu x^\mu} \exp[- \sin s (H x^1 - P_1 t)] e^{-sM}
$$
so
$$
e^{-s M} |\Psi\rangle =  \int d^d x f(x) e^{i \cos s P_\mu x^\mu} \exp[- \sin s (H x^1 - P_1 t)] e^{-sM} {\cal O}(0) e^{-i P_\mu x^\mu} \vac = 
$$
\be
\label{bwcomp}
=\int d^d x f(x) e^{i \cos s P_\mu x^\mu} \exp[- \sin s (H x^1 - P_1 t)]  {\cal O}(0) \vac 
\ee
To get the last line we used $[M, {\cal O}(0)] = 0 $ and $M\vac = P_\mu \vac =0$. 

We see that if the point $x$ is in the wedge $R$, which implies $0<x^1<|t|$, and if we use the spectrum condition $H\geq|P_1|$ for all states, then we have exponential damping in \eqref{bwcomp} for $0<s<\pi$ and the state $\rs$ is in the domain of the operator $e^{-sM}$. Similar results can be derived for states with several insertions of operators in the right wedge \cite{Bisognano:1975ih,Bisognano:1976za}.

It is useful to give another heuristic perspective to this result by returning to the example of free field theory in Rindler space. The operator
\be
\label{cboostf}
e^{-s M} = \exp\left[{-s \int_0^\infty d\omega \, \omega \left(a_\omega^\dagger a_{\omega}- \ta_\omega^\dagger \ta_{\omega} \right)}\right]
\ee
looks like a thermal Boltzmann factor at inverse temperature $s$, where the oscillators in wedge $R$ have positive frequencies, but those in wedge $L$ have negative frequencies. So in general exciting the left oscillators in a state $\rs$ will lead to exponential growth of the norm of $e^{-sM}\rs$.

At first one might think that operators on the right wedge will simply change the occupation level of the right oscillators and since they have positive energy according to \eqref{cboostf} this would lead to a convergent result. This would imply that any state ${\cal A}_R|0\rangle$ should be in the domain of the operator \eqref{cboostf} for any $s>0$. However, we notice that even in the vacuum $|0\rangle$ both left and right oscillators are excited \eqref{runru}, but the contributions from the two sides to \eqref{cboostf} cancel. Now, when acting with operators on the right wedge, we can not only increase the expectation value of $a_\omega^\dagger a_{\omega}$, but we can also decrease it, for example by acting with the annihilation operator $a_\omega$ --- we remind that due to the thermal occupation $\eqref{runru}$ this does not annihilate the state $\vac$. When we lower the right oscillators, the cancellation with the left oscillators in \eqref{cboostf} does not happen, so  effectively we get an increase in the exponent. If this happens for too many modes  the state may no longer lie in the domain of the operator \eqref{cboostf} and this is what leads to the condition $0\leq s \leq \pi$.

To see this in more detail, any operator $A_R$ in the right wedge can be expanded in terms of $a_\omega,a^\dagger_\omega$. We consider a state of the form $\rs = A_R |0\rangle$ and then consider 
the norm of the vector $e^{-s M} \rs$, which is
$$
\langle 0| A_R^\dagger e^{-2s M} A_R |0\rangle
$$
The operator $A_R$ will excite the modes $a^\dagger$ on the right, however this leads to suppression \eqref{cboostf}. It will also decrease the occupation level of the right modes, effectively leading to enhancement since it leaves the thermally populated modes on the left without complete cancellation. However, we notice that as long as $s< \pi$ the thermal suppression of \eqref{runru} will win over the enhancement coming from the factor $e^{-2s M}$. For the case $s=\pi$ we may need to be more careful, but the result is settled by \cite{Bisognano:1975ih,Bisognano:1976za}.

\section{Some estimates}
\label{estimateapp}

Consider a typical pure state $\rs= \sum_i c_i |E_i\rangle$ from the microcanonical ensemble centered around energy $E$ and with energy spread $\delta E$. We assume that the specific state $\rs$ has $\ls H \rs = E_0$. Of course the expectation value $E_0$ is close to $E$, but there may be some difference due to the spread $\delta E$ . Consider the expectation value
\be
\label{toest}
\ls A \hat{H} \rs
\ee
where $\hat{H}=H -E_0$. For simplicity we assume that $A$ is Hermitian. If the state $\rs$ had exact energy $E_0$ then we would have $\hat{H} \rs =0$ and \eqref{toest} would be exactly zero. However $\rs$ has some spread in energy so we expect $\hat{H}\rs$ to be small but nonzero. We want to estimate the size of \eqref{toest} for operators $A$ which belong to the small algebra ${\cal A}$. 

We use the ETH \eqref{ethdef} to write
$$
\ls A \hat{H} \rs = \sum_{i} |c_i|^2 f(E_i) (E_i-E_0)  + \sum_{i,j} c_i^* c_j e^{-S/2} R_{ij} g(E_i,E_j) (E_j-E_0)
$$
The second  term is exponentially small due to cancellations between the random phases $c_i$. Hence we concentrate on the first term
\be
\label{estimateseth}
\ls A \hat{H} \rs = \sum_{i} |c_i|^2 f(E_i) (E_i-E_0)  + O(e^{-S/2})
\ee
From the ETH we expect that $f$ will be a smooth function of the energy $E$. Moreover, in relevant examples, such as the big AdS black hole, observables in the small algebra ${\cal A}$ have the property that at large $S$ their thermal expectation values 
in terms of the temperature is $S$ independent\footnote{We remind that $H$ is not an element of the algebra and that the algebra is closed under commutators with the Hamiltonian: $A\in {\cal A}\rightarrow [H,A]\in {\cal A}$.}, see for example \eqref{btz2point}. This implies that if we consider 
$$
{df \over d E }  = {df \over d\beta} {d\beta \over dE}
$$
the first factor on the RHS is $O(S^0)$ while the second factor if $O(1/S)$. Hence all in all we find
\be
\label{slowf}
{df \over dE} = O(1/S)
\ee
Going back to \eqref{estimateseth}, we expand $f$ around the energy $E$ corresponding to the microcanonical ensemble
$$
\ls A \hat{H} \rs = \sum_{i} |c_i|^2 \left[f(E) + {df \over dE}(E_i-E) + ... \right] (E_i-E_0)  + O(e^{-S/2})
$$
and using \eqref{slowf} and the fact that we selected the spread of the microcanonical $\delta E$ to be at most of $O(T S^0)$, we have
$$
\ls A \hat{H} \rs = f(E) \sum_{i} |c_i|^2 (E_i-E) + O(1/S)  
$$
which implies
$$
|\ls A \hat{H} \rs| \lesssim |f(E)| \delta E + O(1/S)  
$$
$$
|\ls A \hat{H} \rs| \lesssim |\Tr[\rho_m A]| \delta E + O(1/S)  
$$
In the main text we wanted to estimate $\ls A \hat{H} \rs$ for operators $A$ which obey $\ls A \rs = O(1/S)$ on typical states $\rs$ with energy width $\delta E$ at most of order $O(T S^0 )$. The argument above shows that for these operators
\be
\label{finalest}
\ls A \hat{H} \rs = O(1/S)
\ee
\subsection{Details on the change of the energy}
\label{appreden}
We revisit equation \eqref{bla55}
\be
\label{bla56}
\delta E = |c|^2 \lsz V^\dagger [H,V]\rsz + \left[|c|^2\lsz V^\dagger  V H\rsz - \lsz H \rsz \right]  
\ee
and we want to show that the second term is $O(1/S)$. First we rewrite it as
$$
\left[|c|^2\lsz V^\dagger  V H\rsz - \lsz H \rsz \right]= $$
$$= \left[|c|^2\lsz V^\dagger  V \hat{H}\rsz - \lsz \hat{H} \rsz \right]  
-E_0 \left[|c|^2\lsz V^\dagger  V \rsz -1 \right]   
$$
The second term is exactly zero by \eqref{normconbb}. The first term can be written as
\be
\label{jj11}
\lsz \left(|c|^2 V^\dagger V -1\right) \hat{H} \rsz
\ee
This is of the $\ls A \hat{H}\rs$ with $A = |c|^2 V^\dagger V -1$. The thermal expecation value of $A$ is $O(1/S)$, hence we find using \eqref{finalest} that \eqref{jj11} is $O(1/S)$.

\bibliographystyle{JHEP}
\bibliography{references}
\end{document}